\documentclass[a4paper,11pt]{article}
\pdfoutput=1 

\usepackage{amsfonts}
\usepackage{amsmath,bm}
\usepackage{amssymb,braket,dsfont,mathtools}
\usepackage{array}
\usepackage{bbold}
\usepackage[style=numeric-comp,sorting=none,subentry]{biblatex}
\usepackage{booktabs}
\usepackage[makeroom]{cancel}
\usepackage[T1]{fontenc}
\usepackage[margin=1in]{geometry}
\usepackage{graphicx,rotate,multicol,caption}
\usepackage[final]{hyperref}
\hypersetup{
	colorlinks=true,
	linkcolor=blue,
	citecolor=blue,
	filecolor=magenta,
	urlcolor=blue
}
\usepackage[utf8]{inputenc}
\usepackage{mathrsfs}
\usepackage{multirow,tabularx}
\usepackage{parskip}
\usepackage{slashed}
\usepackage{soul}
\usepackage{subcaption}
\usepackage[dvipsnames]{xcolor}

\newcommand{\beq}{\begin{equation}}
\newcommand{\eeq}{\end{equation}}
\newcommand{\bea}{\begin{eqnarray}}
\newcommand{\eea}{\end{eqnarray}}

\newcommand{\nn}{\nonumber}

\newcommand{\eq}[1]{eq.\,\eqref{#1}}
\newcommand{\eqs}[1]{eqs.\,\eqref{#1}}

\newcommand{\keV}{{\rm keV}}
\newcommand{\GeV}{{\rm GeV}}

\newcommand{\Br}[1]{{\rm BR}\left({#1}\right)}

\newcommand{\vep}{\varepsilon}

\newcommand{\oscar}[1]{\textcolor{olive}{#1}}       

\begin{document}
\hfill {\small IFIC/22-12,\quad FTUV-22-0331}\\[0.5cm]

\begin{center}
	{\Large \boldmath \bf Anomaly-free ALP from non-Abelian flavor symmetry} \\[1cm]
	{\large
	C. Han$^a\,$\footnote{hanchch@mail.sysu.edu.cn}, 
	M.L. López-Ibáñez$^{b\,}$\footnote{maloi2@uv.es}, 
	A. Melis$^c\,$\footnote{aurora.melis@uv.es}, 
	O. Vives$^d\,$\footnote{oscar.vives@uv.es}, 
	J.M. Yang$^{e,\, f\,}$\footnote{jmyang@itp.ac.cn}}\\[20pt]
 	{\small
 	$^a$ School of Physics, Sun Yat-Sen University, Guangzhou 510275, China \\[5pt]
 	$^b$ Universidad Politécnica de Madrid, ETSIST, C/Nikola Tesla s/n, 28031 Madrid, Spain. \\[5pt]
 	$^c$ Laboratory of High Energy and Computational Physics, NICPB, R\"avala 10, 10143 Tallinn, Estonia \\[5pt]
 	$^d$ Departament de Física Tèorica, Universitat de València \& IFIC, Universitat
 	    de València \& CSIC, \\ Dr. Moliner 50, E-46100 Burjassot (València), Spain.\\[5pt]
 	$^e$ CAS Key Laboratory of Theoretical Physics, Institute of Theoretical Physics \\
 	    Chinese Academy of Sciences, Beijing 100190, P. R. China. \\[5pt]
 	$^f$ School of Physical Sciences, University of Chinese Academy of Sciences,\\
 	    Beijing 100049, P. R. China}
\end{center}
\vspace*{1cm}

\begin{abstract}
\noindent 
Motivated by the Xenon1T excess in electron-recoil measurements, we investigate the prospects of probing axion-like particles (ALP) in lepton flavor violation experiments. In particular, we identify such ALP as a pseudo-Goldstone from the spontaneous breaking of the flavor symmetries that explain the mixing structure of the Standard Model leptons. We present the case of the flavor symmetries being a non-Abelian $U(2)$ and the ALP originating from its $U(1)$ subgroup, which is anomaly-free with the Standard Model group. We build two explicit realistic examples that reproduce leptonic masses and mixings and show that the ALP which is consistent with  Xenon1T anomaly could be probed by the proposed LFV experiments. 
\end{abstract}

\section{Introduction}
\label{sec:intro}
The structure of fermion masses and mixings remains one of the more mysterious and difficult questions still unanswered in high energy physics. The Standard Model (SM) accommodates the observed flavor structures through the Yukawa couplings of the fermions with the Higgs, but it can not explain their intricate structures or even the existence of three families of fermions. 

The use of flavor symmetries, which generate the Yukawa couplings after spontaneous symmetry breaking, seems to be the most promising path to provide answers to these questions. In flavor models, we extend the SM with a flavor symmetry, under which the SM fermions are charged. The SM Yukawa couplings are forbidden by the symmetry, but higher dimensional operators with additional scalars, the flavons, and fermionic mediators are possible. After spontaneous symmetry breaking, the flavon vevs, normalized by the mediator masses, generate the dimensionless Yukawa couplings in a  {\it Froggatt-Nielsen} mechanism~\cite{Froggatt:1978nt}.
In the literature, there is a large variety of models with different flavor symmetries \cite{Froggatt:1978nt,Leurer:1992wg,Leurer:1993gy,Nir:1993mx,Dine:1993np,Ibanez:1994ig,Pomarol:1995xc,Barbieri:1995uv,Dudas:1995yu,Barbieri:1996ww,Binetruy:1996cs,Choi:1998wc,Ma:1991eg,King:2001uz,Babu:2002dz,Altarelli:2002sg,Ross:2004qn,Chankowski:2005qp,Altarelli:2006kg,deMedeirosVarzielas:2005ax,Luhn:2007sy,Babu:2011mv,Chen:2014wiw,Das:2016czs,Lopez-Ibanez:2017xxw,deMedeirosVarzielas:2017sdv,Petcov:2018snn,Rahat:2018sgs,Lopez-Ibanez:2019rgb,Perez:2019aqq}, either continuous or discrete, Abelian or non-Abelian, that try to explain the structure of Yukawas and neutrino masses. All these flavor symmetries can be gauged or global symmetries but both possibilities have different physical consequences.
In particular, as it is well-known, when the flavor symmetry is global, its spontaneous breaking generates light (pseudo-)Goldstone bosons that can have important effects in the low-energy phenomenology. 

Recently, the Xenon1T experiment reported an excess on electronic recoil events~\cite{XENON:2020rca}, indicating the possible existence of an axion-like particle (ALP) coupling to the electrons~\cite{Takahashi:2020bpq,Han:2020dwo, Takahashi:2020uio}. Then, it would be interesting to identify such ALP as the "flavored" pseudo-Goldstone boson accounting for the mixing structure of lepton sectors. However, to avoid the constraints from astrophysics observations, the global symmetry generating the ALP should be anomaly free with electromagnetic fields, resulting in additional requirements on the structure of this global symmetry. Actually such leptophilic ALP has attracted much attentions in these days~\cite{Calibbi:2020jvd, Choi:2020rgn,Athron:2020maw,Takahashi:2020uio,Chala:2020wvs,Bauer:2020jbp,Chang:2021myh,Darme:2021cxx,Kim:2021eye,Ren:2021prq}.

In a recent paper \cite{Han:2020dwo}, we studied an axion-like particle (ALP) with family-dependent charges as a possible source for the recent Xenon1T excess.
In these models, we have a $U(1)_f$ global symmetry, part of a bigger flavor symmetry, spontaneously broken by the vev of a complex scalar field, $\phi$, whose angular component is identified with an ALP.
The couplings of this $\phi$ field are flavor dependent (in the lepton sector) and thus the ALP will have lepton flavor-changing couplings. These couplings are basically determined by the mixing matrices diagonalizing the charged lepton Yukawas from the flavor basis with diagonal $U(1)_f$ charges.
If this $U(1)_f$ symmetry is the only responsible for the observed hierarchy among the charged-lepton generations, the charged-lepton mixing matrices tend to be CKM-like. Then, ALP flavor changing effects are small and out of reach for the expected sensitivity of proposed experiments like Mu3e or MEGII-fwd. 

However, from the observed PMNS matrix \cite{deSalas:2020pgw, Esteban:2020cvm}, we know that leptonic mixings are large and, in fact, nearly maximal in some sectors. Thus, we can naturally expect large mixings both in the neutrino and charged-lepton mass matrices. Therefore, it is equally legitimate to explore the possibility that these large mixings come mainly from the charged-lepton or from the neutrino sector. In this paper, we follow the large charged-lepton mixing possibility and explore the associated phenomenology in the presence of a "flavored" pseudo-Goldstone boson. As pointed out in \cite{Han:2020dwo}, in this case, we would have larger LFV effects that could be observable in the near-future proposed experiments. We will explore two different realistic flavor models with large charged-lepton mixings, able to reproduce lepton masses and mixings. Both examples are based in a non-Abelian $U(2)_f \,=\, SU(2)_f \,\times\, U(1)_f$ flavor symmetry, where the $U(1)_f$ is global and anomaly-free, with two different assignments for the $U(2)_f$ representations. 

The paper is organized as follows. In the next section, we obtain the ALP couplings to SM leptons and show that they depend only on the charges and charged-lepton mixing matrices. Then we list the constraints and sensitivity from LFV experiments and Xenon1T observations. Section \ref{sec:genmodels} analyzes the required ingredients to have sizeable LFV ALP couplings in  flavor models. In section \ref{subsec:U2f} we build two different realizations of $U(2)_f$, assigning to a doublet the first and second families in \ref{sec:u2f12} and the second and third families in \ref{sec:u2f23}. Finally, in section \ref{sec:conclusions} we present our conclusions.

\section{ALP couplings and phenomenology}
\label{sec:ALP}
In this section, we present the ALP couplings with the SM fermions and the phenomenological constraints that must be taken into account for our analysis.

In general, if we have several flavon scalars charged under the $U(1)_f$ group, the pseudoscalar parts of those fields, $a_i$, mix to produce the physical ALP, $a$, as
\beq
    a ~=~ \sum_i\, \frac{Q_i\, v_i\, a_i}{\sqrt{\sum\, Q^2_j\, v_j^2 }},
\eeq
where $Q_i$ refers to the charge of the flavons under $U(1)_f$ and $v_i$ to the vevs.

Due to the Nambu-Goldstone nature of the ALP field, the interactions between the SM fermions and the ALP are derivative. In this case, they are flavor dependent if $U(1)_f$ couplings are family-dependent, and, in the mass basis, are given by
\beq
    -{\cal L}_{ae} ~=~ i\frac{\partial_\mu a}{2 f_a}\, \bar e_i\, \gamma^\mu \left( V^e_{ij} \,+\, \gamma^5 A^e_{ij} \right)\, e_j,
\eeq
with $f_a\simeq {\mathcal O} (v_i)$ the ALP decay constant and the axial and vector couplings defined as
\bea
    V^e_{ij} & = & \frac{1}{2} \left(U^{e\,\dagger}_R x_R U^e_R \;+\; U^{e\,\dagger}_L x_L U^e_L\right), \label{eq:Ve1} \\
    A^e_{ij} & = & \frac{1}{2} \left(U^{e\,\dagger}_R x_R U^e_R \;-\; U^{e\,\dagger}_L x_L U^e_L\right). \label{eq:Ae1}
\eea

Here the diagonal matrices $x_L$ and $x_R$ are determined by the $U(1)_f$ charges of the corresponding left- (LH) and right-handed (RH) charged leptons:
\beq \label{eq:xL}
    x_L \:=\: {\rm Diag}\left(Q_1,\, Q_2,\, Q_3\right),\hspace{1.cm}
    x_R \:=\: {\rm Diag}\left(q_1,\, q_2,\, q_3\right),
\eeq
and $U^{e}_L$, $U^{e}_R$ are the two rotations that diagonalize the lepton mass matrix
{\beq 
\label{eq:d12_YeDiag}
     U^{ e\, \dagger}_L\; M_e\; U^e_R ~=~ {\rm diag}\left(m_e,\, m_\mu,\, m_\tau\right).
 \eeq}
By using unitarity of $U^e_L$ and $U^e_R$, \eqs{eq:Ve1} and \eqref{eq:Ae1} can be written as:
\bea
    V^e_{ij} & = & \frac{1}{2}  \left[ (q_3+Q_3)\, \delta_{ij} \;+\; (q_1-q_3)\, {U^{e\,*}_R}_{1i} 
        {U^e_R}_{1j} \;+\; (q_2-q_3)\, {U^{e\,*}_R}_{2i} {U^e_R}_{2j} \right. \nn\\
        &&\left. \;+\; (Q_1-Q_3)\,  {U^{e\,*}_L}_{1i} {U^e_L}_{1j} \;+\; (Q_2-Q_3)\,  
        {U^{e\,*}_L}_{2i} {U^e_L}_{2j} \right], \label{eq:Ve2} \\
    A^e_{ij} & = &  \frac{1}{2}  \left[ (q_3-Q_3)\, \delta_{ij} \;+\; (q_1-q_3)\, {U^{e\,*}_R}_{1i} 
        {U^e_R}_{1j} \;+\; (q_2-q_3)\, {U^{e\,*}_R}_{2i} {U^e_R}_{2j} \right. \nn\\
        &&\left. \;-\; (Q_1-Q_3)\,  {U^{e\,*}_L}_{1i} {U^e_L}_{1j} \;-\; (Q_2-Q_3)\,  
        {U^{e\,*}_L}_{2i} {U^e_L}_{2j} \right]. \label{eq:Ae2}
\eea
ALP interactions, at scales much below the symmetry breaking scale, are given by these couplings, $V^e$ and $A^e$. 
From \eqs{eq:Ve2} and \eqref{eq:Ae2}, it can be seen that, for non-equal charges ${\mathcal{O}}(1)$, the size of flavor-changing effects is basically determined by the LH and RH mixing, $U^e_L$ and $U^e_R$.

As discuss at refs.\cite{Bjorkeroth:2018dzu, Calibbi:2020jvd}, flavored ALPs can be searched through LFV processes such as $\ell_j \to \ell_i\, a$, whose branching ratio is given by
\begin{equation} \label{eq:BRljlia}
    \Br{\ell_i\to\ell_j a} \,=\, \frac{m_{\ell_i}^3}{16\pi\Gamma(\ell_j)}\frac{\left|C_{ij}^e\right|^2}{4\, f_a^2} \left( 1-\frac{m_a^2}{m^2_{\ell_i}}\right)^2,
\end{equation}
with $\left|C^e_{ij}\right|^2 = \left|V^e_{ij}\right|^2+\left|A^e_{ij}\right|^2$.
Limits from the LFV decays $\ell_j \to \ell_i \gamma$ and $\ell_j \to 3 \ell_i$ can be imposed over the ALP process in such a way that, once the ALP couplings to leptons are known, they can be translated into bounds on $f_a$.
Present and expected sensitivities on the LFV transitions involving the ALP are collected at Table \ref{tab:LFV}.
The strongest restrictions come from the $\mu\to e\, a$ decay and imply,
\begin{eqnarray}
   f_a^{\text{Jodidio}} &\geq& \left(2.7\times 10^9\,\text{GeV}\right)\left|C_{21}^e\right|, \label{eq:faJod}\\
   f_a^{\text{Mu3e}} &\geq& \left(1.6\times 10^{10}\,\text{GeV}\right)\left|C_{21}^e\right|. \label{eq:faMu3e}
\end{eqnarray}

\begin{table}[t!]
	\centering
	{\renewcommand{\arraystretch}{1.}
	\resizebox{0.55\textwidth}{!}{
	\begin{tabular}{l c l}
    \toprule
    \bf Lepton decay & \bf \quad BR limit \quad\qquad & \bf Experiment \\[2.5pt]
    \midrule
    \multicolumn{3}{c}{Present best limits} \\[2.5pt]
    \midrule
    $\Br{\mu \to e\, a}$ & $< 2.6\cdot 10^{-6}$ & \texttt{Jodidio et al.} \cite{Jodidio:1986mz} \\[2.5pt]
    $\Br{\mu \to e\, a}$ & $< 2.1\cdot 10^{-5}$ & \texttt{TWIST} \cite{TWIST:2014ymv} \\[2.5pt]
    $\Br{\mu \to e\, a\, \gamma}$ & $< 1.1 \cdot 10^{-9}$ & \texttt{Crystal Box} \cite{Bolton:1988af} \\[2.5pt]    
    $\Br{\tau \to e\, a}$ & $< 2.7\cdot 10^{-3}$ & \texttt{ARGUS} \cite{ARGUS:1995bjh} \\[2.5pt]    
    $\Br{\tau \to \mu\, a}$ & $< 4.5\cdot 10^{-3}$ & \texttt{ARGUS} \cite{ARGUS:1995bjh} \\[2.5pt]    
    \midrule
    \multicolumn{3}{c}{Projections of running experiments} \\[2.5pt]
    \midrule
    $\Br{\mu \to e\, a}$ & $< 1.3\cdot 10^{-7}$ & MEGII-fwd \cite{Baldini:2018nnn, Calibbi:2020jvd} \\[2.5pt] 
    $\Br{\tau \to e\, a}$ & $< 8.4\cdot 10^{-6}$ & \texttt{Belle-II} \\[2.5pt]
    $\Br{\tau \to \mu\, a}$ & $< 1.6\cdot 10^{-5}$ & \texttt{Belle-II} \\[2.5pt]    
    \midrule
    \multicolumn{3}{c}{Projections of planned experiments} \\[2.5pt]
    \midrule 
    $\Br{\mu \to e\, a}$ & $< 7.3\cdot 10^{-8}$ & \texttt{Mu3e} \cite{Perrevoort:2018okj} \\[2.5pt]
    \bottomrule
    \end{tabular}}}
\caption{\label{tab:LFV}
        Limits over the axion decay constant from lepton decays.
        \texttt{Belle-II} limits are derived from the simulated result at \texttt{Belle} \cite{Griessinger:2017rpx} by rescaling the luminosity \cite{Calibbi:2020jvd}.}
\end{table}
In addition, there is an interesting proposal at PSI, the MEGII-fwd \cite{Calibbi:2020jvd}. This proposal is based on a detector in the forward direction, where the SM background is suppressed, to collect energetic forward positrons. If we assume perfect $\mu^+$ polarization and with the detector in the forward direction, this decay is only sensitive to $|V^e_{21} + A^e_{21}|^2$ and not to $(V-A)$ couplings \cite{Calibbi:2020jvd}. Therefore, the bound on $f_a$ would be, 
\begin{eqnarray} \label{eq:faMEGII}
   f_a^{\text{MEGII-fwd}} &\geq& \left(1.2\times 10^{10}\,\text{GeV}\right)\left|V_{21}^e + A_{21}^e\right|,
\end{eqnarray}
where the most stringent constraint is obtained in the case of $V_{21}^e=A_{21}^e$.
Moreover, if we require the ALP to explain the Xenon1T excess \cite{XENON:2020rca}, it can only constitute the $\sim 7\%$ of the total dark matter abundance and its coupling to electrons must be\footnote{
In ref.\cite{Takahashi:2020bpq}, eq.(2) and discussion below, it is explained that the Xenon1T excess implies $f_a/q_e \simeq \sqrt{r}\, 10^{10}\, \GeV$, with $r$ the fraction of DM constituted by the ALP.
Then, $g_{ae} \simeq 5\times 10^{-14}/\sqrt{r}$.
In our notation, $g_{ae}$ is the pseudoscalar coupling of the axion with the electron, $P_{11}^e/f_a$, with $P_{ij}^e = (m_i+m_j) A_{ij}^e$.
Then, for $r=0.07$, $A_{11}^e\simeq 2 \times 10^{-13}/(2\, m_e)\, f_a = 10^{-13}\, f_a/m_e$.
As our models fix $A_{11}^e$, a prediction for $f_a$ can be derived as $f_a = A_{11}^e\, m_e/10^{-13}$.
} \cite{Takahashi:2020bpq}:
\begin{equation} \label{eq:Ae11}
    A^e_{11}\simeq 10^{-13}\frac{f_a}{m_e}\,, \hspace{0.5cm} {\rm for}~ m_a\in [2,3]\,\keV.
\end{equation}
Furthermore, for this range of masses, the ALP has to be anomaly-free, $\sum_i Q_i = \sum_i q_i = 0$, so that its interaction with photons is very suppressed and can evade present limits on X-ray emissions.

\section{Building anomaly-free global flavour models}
\label{sec:genmodels}
As seen in the previous section, ALP flavor-changing couplings in the lepton sector are strongly dependent on the unitary matrices diagonalizing the  charged-lepton Yukawa matrix. 
Additionally, the mass of the ALP and its decay constant are fixed if we require it to explain the Xenon1T excess.
The absence of the electromagnetic anomaly to evade X-ray constraints puts an additional restriction on the fermionic charges.
Following the strategy of our previous paper \cite{Han:2020dwo}, we require anomaly cancellation with the SM particle content, without the addition of new fermions, and this leads us to consider flavor dependent charges.  

In our scenario, the flavor symmetries describing fermion masses must include a global $U(1)$ group with non-universal charges and these charges must cancel the electromagnetic anomaly. Therefore, we have to assign different $U(1)$ charges to the three charged-leptons above the scale of flavor symmetry breaking. This implies that the three charged-leptons can not belong to a single triplet of flavor, which would have a single $U(1)$ charge, but there must be three different representations of dimension 1 or, at most, a doublet and a singlet in the left and right-handed sectors. 
Hence, we will consider a $U(1)_{f}$ and a $U(2)_f = SU(2)_f \times U(1)_f$ symmetries as examples of these two situations.

\subsection{A \texorpdfstring{\boldmath $U(1)_f$}{Lg} flavor symmetry}
\label{subsec:genU1f}

$U(1)$ flavor symmetries were used in the first attempts to explain the structure of fermion masses and mixings and they remain a viable option \cite{Froggatt:1978nt,Leurer:1992wg,Leurer:1993gy,Ibanez:1994ig,Pomarol:1995xc,Dudas:1995yu,Binetruy:1996cs,Choi:1998wc,Chankowski:2005qp}.  
However, we must impose the additional requirement of anomaly cancellation with the leptonic charges, $- Q_1 -Q_2 = Q_3$ and $- q_1 -q_2 = q_3$.

If the Yukawa Lagrangian is
\begin{eqnarray} \label{eq:LY}
    \mathcal{L}_{Y} & \supset & Y^e_{ij}\, \bar \ell_{iL}\, \widetilde H_2\, e_{jR} 
        +\; Y^\nu_{ij}\, \bar \ell_{iL}\,  H_2\, \nu_{jR} \;+\; M^\nu_{ij}\, \nu_{iR}\, 
        \nu_{jR}^c,
\end{eqnarray}
we can always use the Higgs charge to make the (3,3) element ${\cal O} (1)$ or the required power, i.e $-Q_3+q_3+q_H \geq 0$, and we take the flavon charge $q=-1$. The required hierarchy imposes then, $(- Q_3) \leq ( -Q_2) \leq (- Q_1)$ and $q_3 \leq q_2 \leq q_1$ with $(- Q_1), q_1 \geq 0$ and $(- Q_3), q_3 \leq 0$ (remember that we impose that the sum of charges must be zero, independently for left and right fields). Now, we can consider, for simplicity, the 2--3 sector, and we have,

\beq
    Y^e/y_\tau ~=~ \begin{pmatrix}
            \vep^{Q_3 - Q_2 + q_2- q_3} & \vep^{Q_3 - Q_2} \\
            \vep^{q_2- q_3}  & 1
            \end{pmatrix} \quad = \quad \begin{pmatrix}
            \vep^{-Q_1 - 2Q_2 + q_1 + 2q_2} & \vep^{ -Q_1 - 2Q_2} \\
            \vep^{q_1 + 2q_2}  & 1
            \end{pmatrix},
\eeq
where, we have suppressed ${\cal O}(1)$ coefficients in all the entries.
Now, we must require that the second eigenvalue, the muon mass, is ${\cal O}(\lambda_c^2)$. 
If we take $\vep = \lambda_c$, we must have $ -Q_1 - 2 Q_2 + q_1 + 2 q_2 = 2$, and the only possibilities would be: i) $ -Q_1 - 2 Q_2 = 1$ and $q_1+2 q_2= 1$, ii) $ -Q_1 -2 Q_2 = 0$ and $q_1+2 q_2= 2$ and iii)  $ - Q_1 - 2 Q_2 = 2$ and $q_1+2 q_2= 0$. 
From here, we can see that only the solution ii) would give large leptonic mixing in the left-handed charged lepton sector.  However, in this case, we have $Q_1 = -2 Q_2$ and $q_1= 2 -2 q_2$. 
Given that $(-Q_1),q_1 \geq (-Q_2), q_2$, we have necessarily $(-Q_2),q_2 \leq 0$. Moreover, if we require $m_e/m_\mu \simeq \lambda_c^2$ ($m_e/m_\mu \simeq \lambda_c^3$), the only possibilities would be $Q_2 = 0$ and $q_2=0$ ($Q_2=0$ and $q_2=-1/3$). 

Notice this structure is not changed if we add the first generation, as we have chosen the charges hierarchical and the first row or the first column can be at most of the same order as the elements in this submatrix.
Therefore they can not change the order of magnitude of these elements. Furthermore, the charges of the first generation are already fixed by this submatrix and the anomaly cancellation condition. The full $3\times 3$ matrix with $Q_2=0$ and $q_2=-1/3$ is
\beq \label{eq:U1genYe}
    Y^e/y_\tau ~=~ \begin{pmatrix}
                    \vep^5 & \vep^2 & 1 \\
                    \vep^5 & \vep^2 & 1 \\ 
                    \vep^5 & \vep^2 & 1
            \end{pmatrix}.
\eeq
As we can see, this matrix gives rise to ${\mathcal O}(1)$ LH mixings, but, in principle, small, CKM-like RH mixing. Neverthless, the large LH mixings are due to $Q_1=Q_2=Q_3$ and therefore, from \eqs{eq:Ve2}--\eqref{eq:Ae2}, left-handed mixings disappear in offdiagonal $V^e$ and $A^e$ ALP couplings. Thus, ALP couplings will depend only on the small RH mixings and therefore will be out of reach for the near future experiments. Unfortunately, the situation is the same in case iii) with ${\mathcal O}(1)$ RH mixings, but equal RH charges and even case i), where we have medium, but still small, LH and RH mixings, can not reach the required size for the near future experiments.

\subsection{A  \texorpdfstring{\boldmath $U(2)_f$}{Lg} flavor symmetry}
\label{subsec:genU2f}
Let us consider now the flavour group $ U(2)_f \,=\, SU(2)_f \,\times\, U(1)_f$ \cite{Barbieri:1995uv,Barbieri:1996ww,Barbieri:1999km,Aranda:1999kc,Blazek:1999hz,Chen:2000fp,Chen:2001pra,Aranda:2001rd,Barbieri:2011ci,Blankenburg:2012nx,Barbieri:2012uh}. 
Here we group the three generations in a doublet and a singlet of $SU(2)_f$ with different $U(1)_f$ charges, giving then rise to FC couplings. 
We consider two possibilities: 

\text{\bf i) 12-doublet case ---} LH and RH fermions of the third generation transform as singlets of $SU(2)_f$ whilst those of the first and second generation belong to doublets,
\beq
     \bar \ell_L \,\equiv\,   \begin{pmatrix}
                        \bar e_L \\
                        \bar \mu_L
                        \end{pmatrix},\qquad 
    \bar \nu_{L} \,\equiv\,   \begin{pmatrix}
                        \bar \nu_{e L} \\
                        \bar \nu_{\mu L}
                        \end{pmatrix},\qquad                   
    \ell_R \,\equiv\,    \begin{pmatrix}
                    e_R \\
                    \mu_R
                    \end{pmatrix}.
\eeq

\text{\bf ii) 23-doublet case ---} LH and RH fermions of the first generation are singlets while those of the second and third generation transform as doublets,
\beq
    \bar \ell_L \,\equiv\,   \begin{pmatrix}
                        \bar \mu_L \\
                        \bar \tau_L
                        \end{pmatrix},\qquad
    \bar \nu_{L} \,\equiv\,   \begin{pmatrix}
                        \bar \nu_{\mu L} \\
                        \bar \nu_{\tau L}
                        \end{pmatrix},\qquad                   
    \ell_R \,\equiv\,    \begin{pmatrix}
                    \mu_R \\
                    \tau_R
                    \end{pmatrix}.
\eeq

Considering $Q$ and $q$ the $U(1)_f$ charges of the LH and RH fields, the following charge matrices determine the ALP couplings to charged leptons in the 12-- and 23--doublet case, respectively:
\beq
    x_L^{(12)} ~=~ {\rm Diag}\left(Q,\, Q,\, -2Q\right), \hspace{1.cm}               x_R^{(12)} ~=~ {\rm Diag}\left(q,\, q,\, -2q\right).
\eeq
\beq
    x_L^{(23)} ~=~ {\rm Diag}\left(-2Q,\, Q,\, Q\right), \hspace{1.cm}              x_R^{(23)} ~=~ {\rm Diag}\left(-2q,\, q,\, q\right).
\eeq

Notice that \eqs{eq:Ve2} and \eqref{eq:Ae2} can be simplified if the universal part is separated from the $x_L$ and $x_R$ charge matrices:
\bea
    V^e_{ij} & = & \frac{1}{2} \left[(q+Q)\, \delta_{ij} \;-\; 3\,q\, U^{e\,*}_{R,\, 
        {\bf s}i}\; U^e_{R,\, {\bf s}j} \;-\; 3\,Q\,  U^{e\,*}_{L,\, {\bf s}i}\;
        U^e_{L,\, {\bf s}j} \right], \label{eq:Veu2} \\
    A^e_{ij} & = & \frac{1}{2} \left[(q-Q) \delta_{ij} \;-\; 3\,q \,U^{e\,*}_{R,\, {\bf s}i}\; U^e_{R,\, {\bf s}j} \;+\; 3\, Q\, U^{e\,*}_{L,\, {\bf s}i}\; U^e_{L,\, {\bf s}j} \right], \label{eq:Aeu2}
\eea
with ${\bf s}=3$ for the 12--doublet case and ${\bf s}=1$ for the 23--doublet case.
From eq.\,\eqref{eq:Aeu2}, the Xenon1T requirement in eq.\,\eqref{eq:Ae11} can be recast as:
\bea \label{eq:faXu2}
    f_a^{\rm X1T} & = & (2.5\times 10^{9}\,\text{GeV})\, \left| \,q \:-\: Q \:+\: 3\, Q\, \left|U^e_{L,\, {\bf s}1}\right|^2 \;-\; 3\, q\, \left|U^e_{R,\, {\bf s}1}\right|^2\, \right|.
\eea

\begin{figure}[t]
    \centering
    \includegraphics[width=0.8\textwidth]{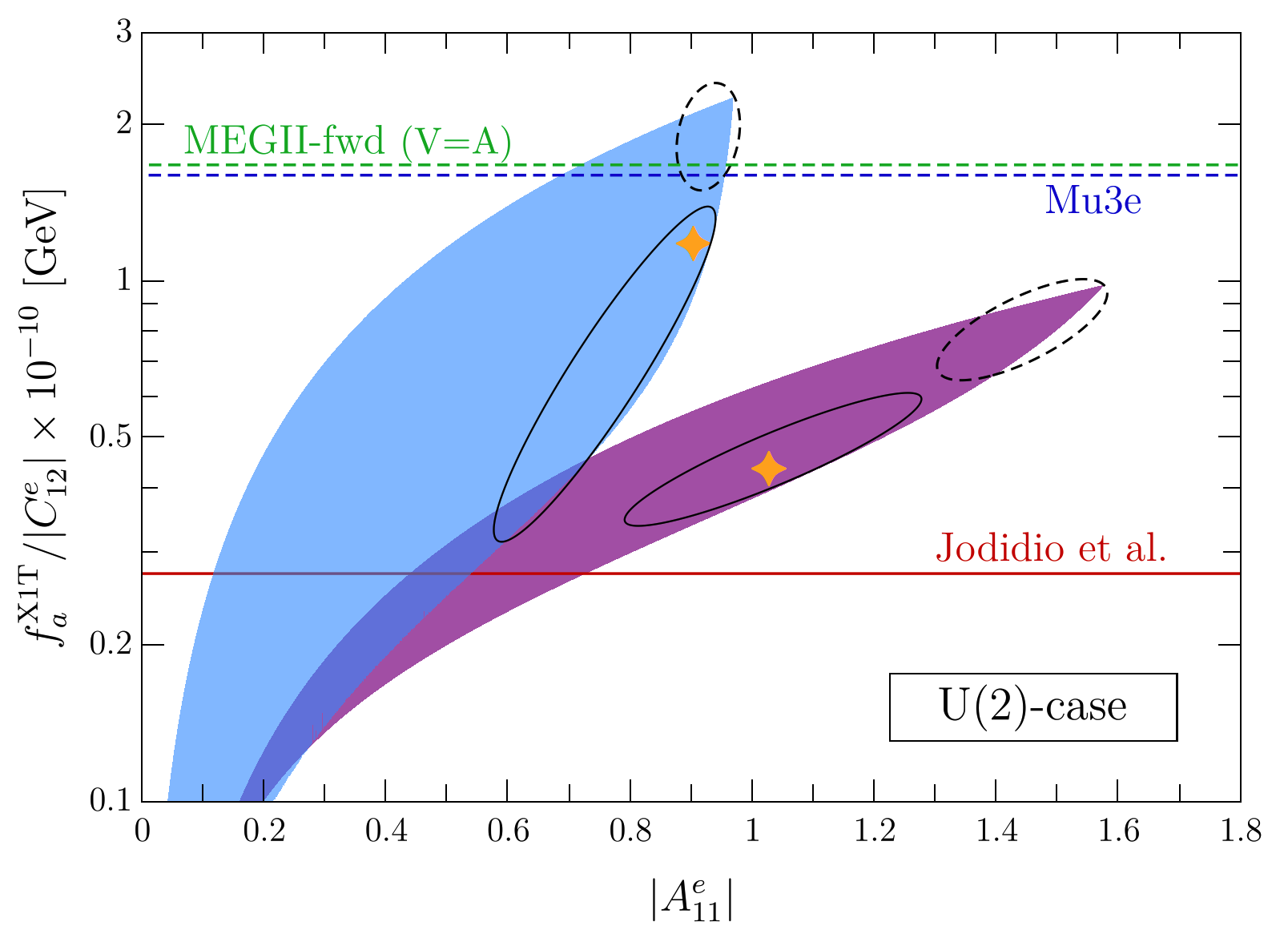}
    \captionsetup{width=\textwidth}
    \caption{\label{fig:fig2}
    Values of the ALP decay constant compatible with the Xenon 1T result assuming a $U(2)_f$ flavor symmetry in the 12-doublet-case (blue region) and 23-doublet-case (violet region) and $Q=-q=1$. 
    The horizontal lines display the bounds in \eqs{eq:faJod}-\eqref{eq:faMu3e}, where the continuous line is the current bound while the dashed lines indicate the limits expected from future experiments. The MEGII-fwd line has been drawn, as a reference, in the limit $V_{21}^e=A_{21}^e$. The full constraint must be obtained in each case from \eq{eq:faMu3e}.
    The LH-mixing is fixed to reproduce the PMNS at the 3$\sigma$ level while the RH-mixing is completely general. 
    The circles identify the regions where the RH-rotation is CKM-like (dashed circles) and PMNS-like (continuous circles). 
    The stars show the predictions corresponding to the benchmark points in Table \ref{tab:d12_coeff} and \ref{tab:d23_coeff}.
    }
\end{figure}

A general view can be obtained from \eqs{eq:Veu2}-\eqref{eq:faXu2} by setting $Q=-q=1$, assuming $U^e_L=U^{3\sigma}_{\rm PMNS}$ and $U^e_R$ totally general. Figure \ref{fig:fig2} displays, in these conditions, the prediction of the ALP decay constant compatible with the Xenon1T result in the 12-doublet (${\bf s} = 3$, blue region) and 23-doublet case (${\bf s} = 1$, violet region).
The dashed and continuous black ellipses correspond to the regions where the RH mixing is CKM- and PMNS-like, respectively.
The different shape of the blue and violet regions can be understood  from \eqs{eq:Veu2}-\eqref{eq:faXu2}. 
For $Q=-q=1$,
\bea
A^{e}_{11} &=& \left(-1 +3/2 \left(\left|U^e_{L,\, { 3 1}}\right|^2 \;+\; \left|U^e_{R,\, {3 1}}\right|^2\right)\right)\,,\\
C^e_{12} &=& \frac{3\sqrt{2}}{2}\left( \left|  U^{e\,*}_{R,\, 
        3 1}\; U^e_{R,\, 3 2}\right|^2 + \left|  U^{e\,*}_{L,\, 
        3 1}\; U^e_{L,\, 3 2}\right|^2 \right)^{1/2} \,,
\eea
in the 12-doublet case or 
\bea
A^e_{11} &=& \left(-1 +3/2 \left(\left|U^e_{L,\, {1 1}}\right|^2 \;+\; \left|U^e_{R,\, {1 1}}\right|^2\right)\right)\,,\\
C^e_{21} &=& \frac{3\sqrt{2}}{2}\left( \left|  U^{e\,*}_{R,\, 
        1 1}\; U^e_{R,\, 1 2}\right|^2 +  \left|  U^{e\,*}_{L,\, 
        1 1}\; U^e_{L,\, 1 2}\right|^2 \right)^{1/2} \,,
\eea
in the 23-doublet case. 

Given that $U^e_L$ is presumed to be PMNS-like, $|U^e_{L,\, { 3 1}}| \sim 0.5$, $|U^e_{L,\, { 3 2}}| \sim 0.6$, $|U^e_{L,\, { 1 1}}| \sim 0.8$ and $|U^e_{L,\, { 1 2}}| \sim 0.55$. 
If $U^e_R$ ranges from small to large values, it is understood that in the 23-doublet case $A^e_{11}$ can reach a factor of two larger than in the 12-doublet case. 
Similarly, as $f_a = 5 \times 10^9~{\rm GeV} |A^e_{11}|$, the slope of the region is given by $ 5 \times 10^9 ~{\rm GeV} /|C^e_{12}|$ and in the 12-doublet case, for small $U^e_R$, the slope is proportional to $1/|C^e_{12}| \simeq 1/0.3 \simeq 3.3 $ while, in the 23-doublet case, it is $1/|C^e_{12}| \simeq 1/0.45 \simeq 2.3$.

The horizontal lines display the bounds in \eqs{eq:faJod}-\eqref{eq:faMu3e}, where the continuous line is the current bound while the dashed lines indicate the limits expected from future experiments. 
It is observed that for both scenarios most of the allowed region by present bounds is testable at future experiments. Only for the 12-doublet case, a small area above the expected Mu3e limit remains unconstrained.

\section{Explicit  \texorpdfstring{\boldmath $U(2)_f = SU(2)_f \times U(1)_f$}{Lg} flavour models}
\label{subsec:U2f}
Let us consider now explicit models with $U(2)_f$ flavor group. These models must reproduce lepton masses and mixings and simultaneously explain the Xenon1T excess. Additionally, we will investigate lepton flavor violation effects in this models and look for points that could be observed at the next generation of experiments. 

As stated before, two possibilities for the leptonic sector are considered: one, where the first and second generation constitute a doublet and the third transforms as singlet, and another, where the role of the first and third generations are exchanged.

In this case, both Higgs doublets, which corresponds to a type-X 2HDM \footnote{In type-X 2HDM, $H_1$ and $H_2$ couple only to quarks and leptons respectively due to an additional $Z_2$ symmetry. Our models are formulated in the large $\tan\beta=v_1/v_2$ regime where the leptophobic doublet can be identified with the SM Higgs in good approximation.
}, transform as singlets under $SU(2)_f$ but have different $U(1)_f$ charges. 

In either case three flavons are added: two transforming as doublets under $SU(2)_f$, $\Phi_1$ and $\Phi_2$, and one as a singlet, $\chi$.
The flavor symmetry is spontaneously broken when the flavons get non-zero vevs in the directions:
\beq
\label{eq:alignment}
    \langle \Phi_1 \rangle ~\propto~ \Lambda\, \begin{pmatrix}
        0 \\
        \vep_1
        \end{pmatrix},\qquad
    \langle \Phi_2\rangle ~\propto~ \Lambda\, \begin{pmatrix}
        \vep_2 \\
        \vep_2
        \end{pmatrix},\qquad
     \langle \chi \rangle ~\propto~ \Lambda\, \vep_\chi.
\eeq
As we show below, the three vevs are of the same order, $\vep_1\sim \vep_2\sim  \vep_\chi$ \footnote{The alignment of these vevs, in particular of $\langle \Phi_2\rangle$, is motivated by the presence of large leptonic mixing. This non-trivial structure can be obtained with an adequate scalar potential, (see for instance \cite{Ross:2004qn,Altarelli:2005yp,Altarelli:2005yx,deMedeirosVarzielas:2005qg,Ding:2011cm})}. Here we note that all the examples below introduce three RH neutrinos that generate the neutrino masses through a type-I see-saw mechanism.

\subsection{The 12-doublet case}
\label{sec:u2f12}
\begin{table}[h!]
	\centering
	{\renewcommand{\arraystretch}{1.5}
	\resizebox{0.9\columnwidth}{!}{
	\begin{tabular}{c c c c c c c c c c c c c }
    \toprule
    \bf Fields & $H_1$ & $H_2$ & $\bar \ell_L, \bar\nu_L$ & $\ell_R$ & $\nu_{eR}$ & $\nu_{\mu R}$ & $\bar \tau_L, \bar\nu_{\tau L}$ & $\tau_R$ & $\nu_{\tau R}$ & $\Phi_1$ & $\Phi_2$ & $\chi$ \\
    \midrule
    \boldmath $\rm SU(2)$ & \bf 1 & \bf 1 & \bf 2 & \bf 2 & \bf 1 & \bf 1 & \bf 1  & \bf 1 & \bf 1 & \bf 2 & \bf 2 & \bf 1 \\
    \boldmath $\rm U(1)$ & 0 & 2/3 & -1/3 & -1/3 & -1/6 & 1/3 & 2/3 & 2/3 & -2/3 & 1 & -2/3 & -1/3 \\
    \bottomrule
\end{tabular}}}
\caption{\label{tab:PartCharg12} \small 
    Particle content and charge assginment under $U(2)_f$ for the 12-doublet case.}
\end{table}

The spectrum and quantum numbers for this version of the $U(2)_f$ model are detailed in table \ref{tab:PartCharg12}.
From the symmetries, the most general set of non-renormalizable operators for charged leptons, at leading order in the flavon vevs, is given by the following couplings with $\Phi_2$:
\bea
    -{\cal L}_e^{(12)} & = & 
                    \frac{c_1^e}{\Lambda^2} \left( \bar\ell_L\, \widetilde\Phi_2 \right) 
                    \left(\ell_R\, \widetilde \Phi_2\right) \widetilde H_2 
                    \;+\; 
                    \frac{c_2^e}{\Lambda^2} \left(\widetilde\Phi_2\, \bar\ell_L \right) \chi\, \tau_R\, \widetilde H_2 \nn \\
                    & + &
                    \frac{c_3^e}{\Lambda^2} \bar\tau_L \left( \widetilde\Phi_2\, \ell_R \right) \chi\, \widetilde H_2
                    \;+\;  
                    \frac{c_4^e}{\Lambda^2}\; \bar\tau_L\, \tau_R\, \chi^2\; \widetilde H_2 
                    \;+\; 
                    {\rm h.c.}\, \label{eq:Lag1},
\eea
where $\widetilde \phi = i \sigma_2 \phi^*$.
The parenthesis in the operators make explicit the {\small $SU(2)_f$} contractions of the fields.
The first order corrections to the terms in eq.\eqref{eq:Lag1} come from the substitution $\widetilde \Phi_2 \rightarrow \Phi_1 \chi$ as well as $\chi \rightarrow \widetilde\Phi_1 \widetilde\Phi_2$:
\bea
    -\delta{\cal L}_e^{(12)} & = &
                    \frac{c^e_5}{\Lambda^3} \left(\bar \ell_L\, \Phi_1\right) \left( \widetilde\Phi_2\, \ell_R \right) \chi\, \widetilde H_2   
                    \;+\; 
                    \frac{c^e_6}{\Lambda^3} \left( \bar\ell_L\, \widetilde\Phi_2 \right) \left(\Phi_1\, \ell_R\right) \chi\, \widetilde H_2 \label{eq:Lag2} \nn \\ 
                    & + &      
                    \frac{c^e_7}{\Lambda^3} \left(\bar \ell_L\, \Phi_1\right) \chi^2\, \tau_R\, \widetilde H_2 
                    \;+\;
                    \frac{c^e_8}{\Lambda^3}\; \bar \tau_L \left(\ell_R\, \Phi_1\right) \chi^2 \widetilde H_2
                    \;+\;
                    {\rm h.c.}\, . \nn
\eea

From eqs.\eqref{eq:Lag1}-\eqref{eq:Lag2}, the resulting mass matrices are
\bea
    M_e^{(12)} & = & \frac{v_{H_2}}{\sqrt{2}}\; \frac{\vep^2_2}{2} \begin{pmatrix}
                c^e_1 & c^e_1 & \sqrt{2}\, c^e_2\, \,\cfrac{\vep_\chi}{\vep_2} \\
                c^e_1 & c^e_1 & \sqrt{2}\, c^e_2\, \,\cfrac{\vep_\chi}{\vep_2} \\
                \sqrt{2}\, c^e_3\, \,\cfrac{\vep_\chi}{\vep_2} & \sqrt{2}\, c^e_3\, \,\cfrac{\vep_\chi}{\vep_2} & c^e_4\, \frac{\vep_\chi^2}{\vep_2^2} \\
                \end{pmatrix}, \label{eq:Ye} \\[10pt]
   \delta M_e^{(12)} & = &  \frac{v_{H_2}}{\sqrt{2}}\; \frac{\vep_1\, \vep_2\, \vep_\chi}{2} \begin{pmatrix}
                c^e_5 + c^e_6 & ~c^e_5~ & ~\sqrt{2}\, c^e_7\, \frac{\vep_\chi}{\vep_2}~ \\[3pt]
                c^e_6 & 0 & 0 \\[3pt]
                ~\sqrt{2}\, c^e_8\, \frac{\vep_\chi}{\vep_2}~ & 0 & \sqrt{2}\, c^e_9 \\
                \end{pmatrix}. \label{eq_dYe}
\eea

In the neutral sector, the Yukawa and Majorana terms that contribute at leading order are\footnote{Equations \eqref{eq:Lnu12} and \eqref{eq:LMnu12} receive NLO corrections from the replacement of flavons discussed above \eq{eq:Lag2}.
However, these NLO operators contribute to exactly the same entries that the leading terms, so they can be safely reabsorbed in the leading {\small ${\cal O}(1)$} coefficients by the redefinition $c_i \to c_i - c_i'\, \vep_1\vep_2/(\sqrt{2}\, \vep_\chi)$.}:
\bea
      -{\cal L}_\nu^{(12)} & = & 
                    \frac{c^\nu_1}{\Lambda} \left(\Phi_2\, \bar\nu_L \right) \nu_{\mu R}\, H_2 
                    \;+\; 
                    \frac{c^\nu_2}{\Lambda^2} \left( \widetilde\Phi_2\, \bar\nu_L \right)\, \chi\, \nu_{\tau R}\, H_2 \nn \\
                    & + &
                    \frac{c^\nu_3}{\Lambda^2} \left( \widetilde\Phi_1 \Phi_2 \right) \bar\nu_{\tau L}\, \nu_{\mu R}\, H_2
                    \;+\;
                    \frac{c^\nu_4}{\Lambda^2} \chi^2\, \bar\nu_{\tau L}\, \nu_{\tau R}\, H_2 \nn \\
                    & + &
                    \frac{c^\nu_5}{\Lambda^2} \left(\widetilde\Phi_1\, \bar\nu_L \right) \chi^\dagger\, \nu_{\mu R}\, H_2 
                    \;+\; 
                    \frac{c^\nu_6}{\Lambda^3} \left(\bar\nu_L\, \widetilde\Phi_1 \right)\, \left(\widetilde\Phi_1 \Phi_2 \right) \nu_{\mu R}\, H_2 \nn \\
                    & + &
                    \frac{c^\nu_7}{\Lambda^3} \left(\bar\nu_L\, \Phi_1 \right) \chi^2\, \nu_{\tau R}\, H_2
                    \; + \; {\rm h.c.}, \label{eq:Lnu12} \\[10pt]   
    {\cal L}_{M}^{(12)} & = & M_0\; \left[\,  
                    \frac{c_8^\nu}{2\Lambda}\, \chi^\dagger\,  \bar\nu^c_{eR}\, \nu_{eR}
                    \;+\;
                    \frac{c^\nu_9}{2\Lambda}\, \chi^\dagger \left(\, \bar\nu^c_{\mu R}\, \nu_{\tau R} + \bar\nu^c_{\tau R}\, \nu_{\mu R} \right) \right. \nn \\
                    & + & 
                    \left. \frac{c_{10}^\nu}{2\Lambda^2}\, \chi^2\, \bar\nu^c_{\mu R}\, \nu_{\mu R} 
                    \;+\;
                    \frac{c_{11}^\nu}{2\Lambda^3}\, \left( \widetilde\Phi_2 \Phi_1 \right) \chi\, \bar\nu^c_{\tau R}\, \nu_{\tau R} 
                    \,\right]
                    \;+\; 
                    {\rm h.c.}\, , \label{eq:LMnu12}
\eea
which produce the following Yukawa and Majorana matrices
\bea
    Y_\nu^{(12)} & = & \frac{\vep_2}{\sqrt{2}} \left(\begin{array}{ccc}
                ~0~ & 
                -c^\nu_1 + c^\nu_6\, \vep_1^2 & 
                 \left(c^\nu_2 + c_7^{\nu}\, \vep_1 \frac{\vep_\chi}{\vep_2}\right) \vep_\chi \\[4pt]
                ~0~ & 
                c^\nu_1 + c^\nu_5\, \vep_1 \frac{\vep_\chi}{\vep_2} & 
                c^\nu_2\, \vep_\chi \\[2pt]
                ~0~ & 
                ~~c^\nu_3\, \vep_1 & 
                \sqrt{2}\, c^\nu_4\, \frac{\vep_\chi}{\vep_2} \vep_\chi
            \end{array}\right)\,, \label{eq:Ynu12} \\[10pt]
    M_\nu^{(12)} & = & M_0\; \vep_\chi  \begin{pmatrix}
                c^{\nu}_8 & 0 & 0 \\
                0  & c_{10}^{\nu}\, \vep_\chi & c_9^{\nu} \\
                0  & c_9^{\nu} & c^{\nu}_{11}\, \frac{\vep_1\vep_2}{\sqrt{2}}  \\
                \end{pmatrix}. \label{eq:Mnu12}
\eea
No coupling involving the RH electron neutrino is present in \eqs{eq:Lnu12} and \eqref{eq:LMnu12}, except for the $\bar\nu^c_{e_R} \nu_{e_R}$ vertex.
This is because the {\small $U(1)_f$} charge of this field has been chosen so that it cannot be compensated by any combination of fields, unless it appears in pairs\footnote{
Unlike $\nu_{\mu_R}$ and $\nu_{\tau_R}$, the $\nu_{e_R}$ charge is an odd multiple of $q_\chi/2$.}.
As a consequence, the zeros appearing in the Yukawa and Majorana matrices of eqs.\eqref{eq:Ynu12} and \eqref{eq:Mnu12} are exact at all orders and lead to one massless active neutrino.
As in this formulation the first and second generation of LH neutrinos are in the same doublet, the natural scenario is having $m_{\nu_3}=0$, which implies an inverted hierarchy. 

The active neutrino mass matrix exhibits the following structure at leading order:
\bea \label{eq:mnu12}
    m_\nu & = & -\frac{v^2_{H_2}}{2}\, Y_\nu\, M^{-1}_\nu\, Y^T_\nu \;=\;
          -\frac{v^2_{H_2}}{2}\, \cfrac{\vep^2_2}{M_0}\, \begin{pmatrix}
                a_1 & 
                ~a_2\, \cfrac{\vep_1 }{\vep_2}\,\vep_\chi~ & 
                a_3\, \cfrac{\vep_\chi}{\vep_2} \\[5pt]
                ~a_2\, \cfrac{\vep_1}{\vep_2}\,\vep_\chi~ & 
                -\, a_1 & 
                -a_3\, \cfrac{\vep_\chi}{\vep_2}  \\[5pt]
                a_3\, \cfrac{\vep_\chi}{\vep_2} & 
                -a_3\, \cfrac{\vep_\chi}{\vep_2} & 
                ~a_4\, \cfrac{\vep_1}{\vep_2}\,\vep_\chi~  \\
                \end{pmatrix} \,.
\eea
with
\beq \label{eq:a1}
    a_1 ~=~ c_1^\nu c_2^\nu, \hspace{1.cm}
    a_2 ~=~ -\frac{c_1^\nu c_7^\nu + c_2^\nu c_5^\nu}{2}, \hspace{1.cm}
    a_3 ~=~ \frac{c_1^\nu c_4^\nu}{\sqrt{2}}, \hspace{1.cm}
    a_4 ~=~ -\sqrt{2}\, c_3^\nu c_4^\nu.
\eeq

The light active neutrinos are Majorana-like, therefore the rotation to the mass basis is just given by the orthogonal transformation
\beq
    U_{\nu_L}^T\, m_\nu\, U_\nu ~=~ {\rm diag}\left(m_{\nu_1},\, m_{\nu_2}, 0\right).
\eeq

\begin{table}[t!]
	\centering
	{\renewcommand{\arraystretch}{1.3}
	\resizebox{0.75\columnwidth}{!}{
	\begin{tabular}{c | c c c c c c c c c c c c c c c c c c c c c}
    \toprule
    $\chi^2_{3\sigma}$ & $c^e_2$ & $c^e_3$ & $c^e_4$ & $c^e_5$ & $c^e_6$ & $c^e_7$ & $c^e_8$ & $\epsilon_1$ & $\epsilon_2$ & $\epsilon_\chi$ \\
    2.16 & -1.30 & 0.93 & -1.63 & -1.43 & -0.82 & -2.00 & -1.02 & 0.05 & 0.17 & 0.15 \\
    \midrule
    F.T. & $c^\nu_1$ & $c^\nu_2$ & $c^\nu_3$ & $c^\nu_4$ & $c^\nu_5$ & $c^\nu_6$ & $c^\nu_7$ & $c^\nu_{10}$ & $c^\nu_{11}$ \\
    8.30 & -0.51 & -3.50 & -2.36 & -0.50 & -1.02 & 1.29 & -1.46 & 0.54 & -0.43 \\
    \bottomrule
\end{tabular}}}
\caption{\label{tab:d12_coeff} \small 
   Values of the model parameters at the benchmark point with $c^e_1,\, c^\nu_8,\, c^\nu_9 = 1.00$.}
\end{table}
\begin{table}[t!]
    \vspace{0.25cm}
	\centering
	{\renewcommand{\arraystretch}{1.3}
	\resizebox{\columnwidth}{!}{
	\begin{tabular}{r | c c c c c c c c}
    \toprule
    & $\theta_{12}^{\rm IH}$ (º) & $\theta_{23}^{\rm IH}$ (º) & $\theta_{13}^{\rm IH}$ (º) &
    $m_e/m_\tau$ & $m_\mu/m_\tau$ & $\Delta m_{21}^2/\Delta m_{23}^2$ \\
    \midrule
    b.p. : & 32.00 & 52.1 & 8.61 & $2.88 \times 10^{-4}$ & $0.059$ & $-0.030$ \\
    \midrule
    exp. : & 33.45 & 49.5 & 8.60 & $2.88\times 10^{-4}$ & $0.059$ & $-0.030$ & $-$ & $-$\\
    $3\sigma$ : & $31.27-35.87$ & $39.8-52.1$ & $8.24-8.98$ & $\left(2.79-2.96\right) \times 10^{-4}$ & $0.0577-0.0612$ & $-\left(0.027-0.033\right)$ \\
    \bottomrule
\end{tabular}}}
\caption{\label{tab:d12_obs} \small
   Value of the measured observables from the \href{www.nu-fit.org}{NuFIT 5.1 (2021)} global fit \cite{Esteban:2020cvm} and predictions of our model at the benchmark point.
   }
\end{table}

A fit is performed in order to fix the values of the free parameters that best reproduce the observed masses and mixing angles in the leptonic sector.
More details about the cost function and how the fit is done can be found in ref. \cite{Fedele:2020fvh}.
Tables \ref{tab:d12_coeff} and \ref{tab:d12_obs} show the coefficients of the model and estimations for the observables at the benchmark point.
Notice that the PMNS angles are mostly given by the rotations of the charged leptons, being the neutrino contribution totally negligible for the 1-2 and 1-3 sectors and a correction of the $20\%$ in the 2-3 sector.
The fit provides the correct ratio of masses between different generations.
From its predictions in table \ref{tab:d12_obs}, the value of the $H_2$ vev and the Majorana mass scale can be determined by setting the absolute value of the masses.
It renders {\small $v_{H_2}=0.17\, v_H$}, with {\small $v_H=246$} GeV the SM Higgs vev, and {\small $M_0=9.04\times 10^{11}$} GeV.

Regarding the ALP phenomenology, the benchmark point of the model produces a decay constant equal to
\beq
    f_a = 1.3\times 10^9~ {\rm GeV}.
\eeq
In figure \ref{fig:fig2}, the position of our benchmark point in the (scaled) ALP parameter space is marked with a yellow diamond in the blue area\footnote{Notice that the blue and purple regions are computed considering $Q=-q=1$ while the charges of our model are $Q=-q=1/3$}. 
As can be seen, an axial coupling to electrons of {\small $|A_{11}|=0.3$} is generated while 
its flavor-violating couplings in the 1-2 sector make feasible to prove its presence through the $\mu \to e\, a$ decay at Mu3e.
However, it is not detectable at MEGII-fwd which can only test up to {\small $f_a/|C_{21}^e| = 8.9\times 10^8 $} GeV for couplings like the ones derived in this scenario {\small ($A_{21} \simeq -0.9\, V_{21}$)}.

\subsection{The 23-doublet case}
\label{sec:u2f23}
\begin{table}[h!]
	\centering
	{\renewcommand{\arraystretch}{1.5}
	\resizebox{0.9\columnwidth}{!}{
	\begin{tabular}{c c c c c c c c c c c c c}
    \toprule
    \bf Fields & $H_1$ & $H_2$ & $\bar e_L,\bar \nu_{eL} $ & $e_R$ & $\nu_{eR}$ & $\nu_{\mu R}$ & $\nu_{\tau R}$& $\bar \ell_L,\bar \nu_L $ & $\ell_R$ & $\Phi_1$ & $\Phi_2$ & $\chi$ \\
    \midrule
    \boldmath $\rm SU(2)$ & \bf 1 & \bf 1 & \bf 1 & \bf 1 & \bf 1 & \bf 1 & \bf 1 & \bf 2 & \bf 2 & \bf 2 & \boldmath $ 2$ & \bf 1 \\
    \boldmath $\rm U(1)$ & 0 & 2/3 & 2/3 & 2/3 & -1/12 & 1/2 & -2/3 & -1/3 & -1/3 & 5/6 & -2/3 & -1/6 \\
    \bottomrule
\end{tabular}}}
\caption{\label{tab:PartCharg23} \small 
    Particle content and charge assignment under $U(2)_f$ for the 23-doublet case.}
\end{table}

This version of the $U(2)_f$ model is very similar to the previous 12-doublet case but with $e_L,\, e_R$ the singlets under $SU(2)_f$.
The charges of $\Phi_1$ and $\chi$ under the $U(1)_f$ symmetry are also slightly modified, as can  be seen in table \ref{tab:PartCharg23}.

The leading operators in the charged sector are exactly those in \eqs{eq:Lag1} and \eqref{eq:Lag2} but making the replacements:
\beq
    \tau_R \to e_R,\qquad \tau_L \to e_L,\qquad \chi \to \chi^2.
\eeq

In view of this, the mass matrix for the charged leptons can then be precisely obtained as $M^{(12)}_e=P^T_{231}.\, M^{(23)}_e(\vep_\chi\rightarrow \vep^2_\chi).P_{231}$, where $P_{231}$ is the $231$ column permutation matrix of the identity.
The resulting charged lepton mass matrix and its first order correction read as
\bea
\label{eq:Me23}
    M^{(23)}_e &=& \frac{v_{H_2}}{\sqrt{2}} 
                \frac{\vep^2_2}{2} \begin{pmatrix}
                2\, c^e_4\, \frac{\vep_\chi^2}{\vep_2^2}\,\vep_\chi^2 & \sqrt{2}\,c^e_2\, \cfrac{\vep_\chi}{\vep_2}\,\vep_\chi & \sqrt{2}\, c^e_2\, \,\cfrac{\vep_\chi}{\vep_2}\,\vep_\chi \\
                \sqrt{2}\, c^e_3\, \cfrac{\vep_\chi}{\vep_2}\,\vep_\chi & c^e_1 & c^e_1 \\
                \sqrt{2}\, c^e_3\, \cfrac{\vep_\chi}{\vep_2}\,\vep_\chi & c^e_1 & c^e_1 \\
                \end{pmatrix},\\[10pt]
   \delta M^{(23)}_e &=&  
                \frac{\vep_1\vep_2\vep_\chi}{2} \begin{pmatrix}
                0 & \sqrt{2} \, c^e_7\, \frac{\vep_\chi}{\vep_2}\,\vep_\chi & 0 \\[5pt]
                \sqrt{2} \,c^e_8\,   \frac{\vep_\chi}{\vep_2}\,\vep_\chi  & c^e_5+c^e_6 
                 + c^e_9\,\frac{\vep_1}{\vep_2}\vep_\chi\, 
                &  c^e_5 \\[5pt]
               0 & c^e_6 & 0 \\
                \end{pmatrix}.
\eea
At variance with the 12-doublet case, in this model, given that the LH neutrinos of the second and third generations belong to the same doublet, a scenario with normal hierarchy for the neutrino mass spectrum, $m_{\nu 1}\ll m_{\nu 2}< m_{\nu 3} $, is the natural prediction. 
With the charge assignment in Table \ref{tab:PartCharg23}, in the neutral sector the following leading order Majorana and Yukawa terms can be written in the Lagrangian
\bea
    -{\cal L}^{(23)}_\nu & = & 
                    \frac{c^\nu_1}{\Lambda} \left( \widetilde\Phi_1\, \bar\nu_L \right)\, \nu_{\mu R}\, H_2 
                    \;+\; 
                    \frac{c^\nu_2}{\Lambda^3} \left( \widetilde\Phi_2\, \bar\nu_L \right)\, \chi^2\, \nu_{e R}\, H_2 \nn\\
                     & + &
                    \frac{c^\nu_3}{\Lambda^4} \left( \widetilde\Phi_1 \Phi_2 \right)\chi^2 \,\bar\nu_{e L}\, \nu_{\mu R}\, H_2 
                    \;+\; 
                    \frac{c^\nu_4}{\Lambda^4} \chi^4\, \bar\nu_{e L}\, \nu_{e R}\, H_2 \label{eq:LYnuLO} \\
                    & + &
                     \frac{c^\nu_5}{\Lambda^2} \left(\Phi_2\, \bar\nu_L \right) \chi\,\nu_{\mu R}\, H_2
                    \;+\; 
                    \frac{c^\nu_6}{\Lambda^4} \left( \Phi_1\, \bar\nu_L \right)\, \chi^3\, \nu_{e R}\, H_2
                    \;+\; {\rm h.c.}, \nn \\[5pt]
    {\cal L}^{(23)}_{M} & = &  
                    M_{0}\,\chi^\dagger\,  \left[ \cfrac{c^\nu_7}{\Lambda} \left( \bar\nu^c_{\mu R}\, \nu_{e R} + \bar\nu^c_{e R}\, \nu_{\mu R} \right) 
                    + 
                    \cfrac{c^\nu_8}{\Lambda}\,\bar\nu^c_{\tau R}\nu_{\tau R}\right] \;+\; {\rm h.c.}.
\eea
Similar to the previous case, we chose the $U(1)_f$ charge of the RH tau neutrino to be a fraction of the smallest flavon charge: $q_{\nu_{eR}}=q_\chi/2$, hence it is never compensated by the considered fields.
Thus, by construction, in $\mathcal{L}^{(23)}_\nu$ there are no couplings involving the RH tau neutrino.
The resulting Yukawa and Majorana mass matrices display the following textures
\bea
    Y^{(23)}_\nu & = &   
                \cfrac{\vep_1}{\sqrt{2}} \begin{pmatrix}
               \sqrt{2}\,c^\nu_4 \frac{\vep_\chi}{\vep_1}\,\vep^3_\chi & c^\nu_3\,\vep_2\,\vep^2_\chi & 0 \\[5pt]
               c^\nu_2\, \frac{\vep_2}{\vep_1}\,\vep^2_\chi - c^\nu_6\,\vep_\chi^3 & c^\nu_5 \frac{\vep_2}{\vep_1}\,\vep_\chi & 0 \\[5pt]
               c^\nu_2\, \frac{\vep_2}{\vep_1}\,\vep^2_\chi & c^\nu_1-c^\nu_5 \frac{\vep_2}{\vep_1}\,\vep_\chi  & 0 \\
                \end{pmatrix}, \label{eq:Ynu23}\\[10pt]
    M_\nu^{(23)} & = & M_0\; \vep_\chi  \begin{pmatrix}
                0 & c^\nu_7 & 0 \\
                c^\nu_7  & 0 & 0 \\
                0  & 0 & c^\nu_8  \\
                \end{pmatrix}. \label{eq:Mnu23}
\eea
Once more, the zeros in the third column of \eq{eq:Ynu23} and the third column and row of \eq{eq:Mnu23} are exact at all orders and lead to a simplified scenario with one massless neutrino. In particular, the simplification of $m_{\nu 1}=0$, implies $m_{\nu 2}=\sqrt{\Delta m^2_{21}}$ and $m_{\nu 3}=\sqrt{\Delta m^2_{31}}$.  

The effective neutrino mass matrix after the seesaw exhibits a the following structure:
\bea
    m^{(23)}_\nu &=& -\frac{v^2_{H_2}}{2} \cfrac{\vep_1\,\vep_2\,\vep_\chi}{M_0}
    \begin{pmatrix}
      2\,a_6\,\vep^4_\chi &  a_5 \,\cfrac{\vep_\chi}{\vep_1}\,\vep^2_\chi &  a_4\,\cfrac{\vep_\chi}{\vep_2}\vep_\chi\\[5pt]
      a_5 \,\cfrac{\vep_\chi}{\vep_1}\,\vep^2_\chi & 2\,a_2\frac{\vep_2}{\vep_1}\,\vep_\chi & a_1-a_3\cfrac{\vep_1}{\vep_2}\,\vep_\chi\\[5pt]
      a_4\,\cfrac{\vep_\chi}{\vep_2}\vep_\chi & a_1-a_3\cfrac{\vep_1}{\vep_2}\,\vep_\chi &  2\,a_1\\
    \end{pmatrix}\,,
\eea
 where the effective parameters are related with the parameters of the lagrangian as
\bea
   a_1=c^\nu_1c^\nu_2,\quad a_2=c^\nu_2 c^\nu_5 ,\quad a_3=c^\nu_1 c^\nu_6,\quad a_4=\sqrt{2}\,c^\nu_1 c^\nu_4,\quad a_5=\sqrt{2}\,c^\nu_4 c^\nu_5,\quad a_6=\sqrt{2}\,c^\nu_3 c^\nu_4\,.
\eea

\begin{table}[t!]
	\centering
	{\renewcommand{\arraystretch}{1.3}
	\resizebox{0.78\columnwidth}{!}{
	\begin{tabular}{c| c c c c c c c c c c c c c c c c c c c c c}
    \toprule
    $\chi^2_{3\sigma}$ & $c^e_1$ & $c^e_2$ & $c^e_3$ & $c^e_4$ & $c^e_5$ & $c^e_6$ & $c^e_7$ & $c^e_8$ & $c^e_9$ \\
     3.69 & 1.649 & -0.451 & -1.74761 & -0.728 & -1.883 & 2.152 & 0.863 & -2.343 & 2.914 \\
      \midrule
     F.T. & $c^\nu_1$ & $c^\nu_2$ & $c^\nu_3$ & $c^\nu_4$ & $c^\nu_5$ & $c^\nu_6$ & $\vep_1$ &  $\vep_2$ & $\vep_\chi$\\
    22  & -0.509 & 0.844 & -1.393 & -1.041 & 2.572 & 2.775 & 0.371 & 0.389 & 0.246 \\ 
      
    \bottomrule
\end{tabular}}}
\caption{\label{tab:d23_coeff} \small 
   Values of the model parameters at the benchmark point with $c^\nu_7=c^\nu_8=1$.}
\end{table}
\begin{table}[t!]
	\centering
	{\renewcommand{\arraystretch}{1.3}
	\resizebox{0.95\columnwidth}{!}{
	\begin{tabular}{c| c c c c c c c }
    \toprule
     & $\theta^{\rm NH}_{12}$\,(º) & $\theta^{\rm NH}_{23}$\,(º) & $\theta^{\rm NH}_{13}$\,(º) &
    $m_e/m_\tau$ & $m_\mu/m_\tau$ & $\Delta m^2_{21}/\Delta m^2_{31}$\\
    \midrule
     b.f. :& 33.56 & 48.0 & 8.55 & $2.88 \times 10^{-4}$ & 0.059 & 0.029\\ 
     \midrule
     exp. :& 33.44 & 49.2 & 8.57 &  $2.88 \times 10^{-4}$ & 0.059 & 0.029 \\
     $3\sigma$ :& $31.27-35.86$ & $39.5-52.0$ & $8.20-8.97$ & $2.79-2.96 \times 10^{-4}$ & 0.0577-0.0612 &  0.027-0.033\\ 
    \bottomrule
\end{tabular}}}
\caption{\label{tab:d23_obs} \small
   Values of the measured observables from the \href{www.nu-fit.org}{NuFIT 5.1 (2021)} global fit \cite{Esteban:2020cvm} and values obtained from the 23-model at the benchmark point. The absolute mass values can be obtained with $v_{H_2}= v_H /26$ and $M_0\simeq 3.17 \times 10^{10}$ GeV. }
\end{table}
In Table \ref{tab:d23_coeff} and \ref{tab:d23_obs}, the values of the coefficients and observables at the benchmark point are presented. From Table \ref{tab:d23_obs} we see that the PMNS is correctly reproduced even at the $1\sigma$ level, being the neutrino contribution 17\% for the $\theta_{23}$ angle, 50\% for the $\theta_{12}$ angle, and 80\% for the $\theta_{13}$ angle. From the Table \ref{tab:d23_obs}, the values of the $H_2$ vev and the Majorana mass scale $M_0$ can be determined by imposing to reproduce the absolute values of $m_\tau$ and $\Delta m^2_{31}$, then we have $v_{H_2}=v_H/26$ and $M_0=3.17 \times 10^{10}$ GeV.
The prediction for the axial ALP coupling to electron is $|A^e_{11}|=0.342$. Consequently, from eq.(\ref{eq:Ae11}), the prediction for the ALP decay constant compatible with the Xenon 1T result reads
\begin{equation}
 f^{\text{X1T}}_a = 1.75 \times 10^9\, \text{GeV}.
\end{equation}
From the requirement of large mixings to come from the charged lepton sector, we obtain large LFV axion couplings with $|A^e_{21}|\gg |V^e_{21}|$, in particular $|C^e_{21}|=0.409$ and $|V^e_{21}+A^e_{21}|=0.484$. With these values the bounds in eq.(\ref{eq:faJod}-\ref{eq:faMEGII}) translates into: $f^{\text{Jodidio}}_a\geq 1.10\times 10^9$ GeV, $f^{\text{Mu3e}}_a\geq 6.5\times 10^9$ GeV and $f^{\text{MEGII-fwd}}_a\geq 5.8\times 10^9$ GeV. So, while the point evades the current limit imposed by the Jodidio et al. experiment, it lies within a region that will be fully tested by the future Mu3e and MEGII-fwd experiments, as it is displayed in Figure \ref{fig:fig2} (yellow diamond inside the violet region).
\section{Conclusions}
\label{sec:conclusions}
Motivated by the Xenon1T excess in electron-recoil measurements, we studied the possible lepton flavor violation effects from an ALP explaining this excess and originating from the spontaneous breaking of a flavor group. This flavor group is built to explain the masses and mixing patterns in the lepton sector. 

A priory, we have two options for flavor symmetries with a non-anomalous $U(1)$ subgroup and flavor dependent charges, $U(1)$ or $U(2)$. We have seen that a plain $U(1)$ symmetry can not provide sizeable flavor-changing couplings reachable in near future experiments. Therefore, we have concentrated in the flavor group $U(2)_f=SU(2)_f\times U(1)_f$. We considered two scenarios, where the first-second or second-third generations of leptons are grouped in a doublet representation of $SU(2)_f$. In these models, we require the symmetry to reproduce correctly lepton masses and mixing with $\mathcal{O}(1)$ coefficients and sizeable mixings in the charged lepton sector. We find that, in both scenarios, the ALP consistent with the Xenon1T anomaly could be probed by future LFV experiments. Interestingly, we find that for the 1-2-doublet case, an inverted hierarchy of neutrino masses is preferred while for the 2-3-doublet case, a normal hierarchy of the neutrino masses is more likely. Therefore, if LFV experiments find a positive signal from ALP interactions, future neutrino experiments could help to disentangle these two models.

\section*{Acknowledgments}
We acknowledge the collaboration of L. Calibbi in the early stages of this work. C. H. is supported by the Guangzhou Basic and Applied Basic Research Foundation under Grant No. 202102020885, and the Sun Yat-Sen University Science Foundation.
MLLI, AM and OV acknowledge support from Spanish AEI-MICINN, PID2020-113334GB- I00/AEI/10.13039/501100011033.
OV acknowledges partial support from the “Generalitat Valenciana” grant PROMETEO2017-033.  A.M. was supported
by the Estonian Research Council grant MOBTT86 “Probing the Higgs sector at the LHC
and beyond”.
JMY acknowledges funding from the National Natural Science Foundation of China (NNSFC) under grant Nos.12075300 and 11821505, from Peng-Huan-Wu Theoretical Physics Innovation Center (12047503), from the CAS Center for Excellence in Particle Physics (CCEPP), and by a Key R\&D Program of Ministry of Science and Technology of China
under number 2017YFA0402204, by the Key Research Program of the Chinese 
Academy of Sciences, grant No. XDPB15.

\printbibliography

@article{Darme:2021cxx,
    author = "Darm\'e, Luc and Nardi, Enrico",
    title = "{Exact accidental U(1) symmetries for the axion}",
    eprint = "2102.05055",
    archivePrefix = "arXiv",
    primaryClass = "hep-ph",
    doi = "10.1103/PhysRevD.104.055013",
    journal = "Phys. Rev. D",
    volume = "104",
    number = "5",
    pages = "055013",
    year = "2021"
}

@article{Bauer:2020jbp,
    author = "Bauer, Martin and Neubert, Matthias and Renner, Sophie and Schnubel, Marvin and Thamm, Andrea",
    title = "{The Low-Energy Effective Theory of Axions and ALPs}",
    eprint = "2012.12272",
    archivePrefix = "arXiv",
    primaryClass = "hep-ph",
    reportNumber = "IPPP/20/69, MITP/20-070 SISSA 30/2020/FISI, ZH-TH-47/20",
    doi = "10.1007/JHEP04(2021)063",
    journal = "JHEP",
    volume = "04",
    pages = "063",
    year = "2021"
}

@article{Chala:2020wvs,
    author = "Chala, Mikael and Guedes, Guilherme and Ramos, Maria and Santiago, Jose",
    title = "{Running in the ALPs}",
    eprint = "2012.09017",
    archivePrefix = "arXiv",
    primaryClass = "hep-ph",
    doi = "10.1140/epjc/s10052-021-08968-2",
    journal = "Eur. Phys. J. C",
    volume = "81",
    number = "2",
    pages = "181",
    year = "2021"
}

@article{Athron:2020maw,
    author = "Athron, Peter and others",
    title = "{Global fits of axion-like particles to XENON1T and astrophysical data}",
    eprint = "2007.05517",
    archivePrefix = "arXiv",
    primaryClass = "astro-ph.CO",
    reportNumber = "CP3-20-36, TTK-20-21, CoEPP-MN-20-5, ADP-20-22/T1132,
  gambit-physics-2020",
    doi = "10.1007/JHEP05(2021)159",
    journal = "JHEP",
    volume = "05",
    pages = "159",
    year = "2021"
}

@article{Perez:2019aqq,
    author = "P\'erez, M. Jay and Rahat, Moinul Hossain and Ramond, Pierre and Stuart, Alexander J. and Xu, Bin",
    title = "{Stitching an asymmetric texture with $\mathcal{T}_{13} \times \mathcal{Z}_5$ family symmetry}",
    eprint = "1907.10698",
    archivePrefix = "arXiv",
    primaryClass = "hep-ph",
    doi = "10.1103/PhysRevD.100.075008",
    journal = "Phys. Rev. D",
    volume = "100",
    number = "7",
    pages = "075008",
    year = "2019"
}

@article{Rahat:2018sgs,
    author = "Rahat, Moinul Hossain and Ramond, Pierre and Xu, Bin",
    title = "{Asymmetric tribimaximal texture}",
    eprint = "1805.10684",
    archivePrefix = "arXiv",
    primaryClass = "hep-ph",
    doi = "10.1103/PhysRevD.98.055030",
    journal = "Phys. Rev. D",
    volume = "98",
    number = "5",
    pages = "055030",
    year = "2018"
}

@article{Barbieri:2012uh,
    author = "Barbieri, Riccardo and Buttazzo, Dario and Sala, Filippo and Straub, David M.",
    title = "{Flavour physics from an approximate $U(2)^3$ symmetry}",
    eprint = "1203.4218",
    archivePrefix = "arXiv",
    primaryClass = "hep-ph",
    doi = "10.1007/JHEP07(2012)181",
    journal = "JHEP",
    volume = "07",
    pages = "181",
    year = "2012"
}

@article{Blankenburg:2012nx,
    author = "Blankenburg, Gianluca and Isidori, Gino and Jones-Perez, Joel",
    title = "{Neutrino Masses and LFV from Minimal Breaking of $U(3)^5$ and $U(2)^5$ flavor Symmetries}",
    eprint = "1204.0688",
    archivePrefix = "arXiv",
    primaryClass = "hep-ph",
    reportNumber = "CERN-PH-TH-2012-082",
    doi = "10.1140/epjc/s10052-012-2126-7",
    journal = "Eur. Phys. J. C",
    volume = "72",
    pages = "2126",
    year = "2012"
}

@article{Barbieri:2011ci,
    author = "Barbieri, Riccardo and Isidori, Gino and Jones-Perez, Joel and Lodone, Paolo and Straub, David M.",
    title = "{$U(2)$ and Minimal Flavour Violation in Supersymmetry}",
    eprint = "1105.2296",
    archivePrefix = "arXiv",
    primaryClass = "hep-ph",
    doi = "10.1140/epjc/s10052-011-1725-z",
    journal = "Eur. Phys. J. C",
    volume = "71",
    pages = "1725",
    year = "2011"
}

@article{Aranda:1999kc,
    author = "Aranda, Alfredo and Carone, Christopher D. and Lebed, Richard F.",
    title = "{U(2) flavor physics without U(2) symmetry}",
    eprint = "hep-ph/9910392",
    archivePrefix = "arXiv",
    reportNumber = "WM-99-117, JLAB-THY-99-31",
    doi = "10.1016/S0370-2693(99)01497-5",
    journal = "Phys. Lett. B",
    volume = "474",
    pages = "170--176",
    year = "2000"
}

@article{Blazek:1999hz,
    author = "Blazek, T. and Raby, S. and Tobe, K.",
    title = "{Neutrino oscillations in an SO(10) SUSY GUT with U(2) x U(1)**n family symmetry}",
    eprint = "hep-ph/9912482",
    archivePrefix = "arXiv",
    reportNumber = "OHSTPY-HEP-T-99-017, NUHEP-TH-99-77",
    doi = "10.1103/PhysRevD.62.055001",
    journal = "Phys. Rev. D",
    volume = "62",
    pages = "055001",
    year = "2000"
}

@article{Chen:2000fp,
    author = "Chen, Mu-Chun and Mahanthappa, K. T.",
    title = "{From CKM matrix to MNS matrix: A Model based on supersymmetric SO(10) x U(2)(F) symmetry}",
    eprint = "hep-ph/0005292",
    archivePrefix = "arXiv",
    reportNumber = "COLO-HEP-445",
    doi = "10.1103/PhysRevD.62.113007",
    journal = "Phys. Rev. D",
    volume = "62",
    pages = "113007",
    year = "2000"
}

@article{Chen:2001pra,
    author = "Chen, Mu-Chun and Mahanthappa, K. T.",
    title = "{CP violation in a supersymmetric SO(10) x U(2)(F) model}",
    eprint = "hep-ph/0106093",
    archivePrefix = "arXiv",
    reportNumber = "COLO-HEP-466",
    doi = "10.1103/PhysRevD.65.053010",
    journal = "Phys. Rev. D",
    volume = "65",
    pages = "053010",
    year = "2002"
}

@article{Aranda:2001rd,
    author = "Aranda, Alfredo and Carone, Christopher D. and Meade, Patrick",
    title = "{U(2) like flavor symmetries and approximate bimaximal neutrino mixing}",
    eprint = "hep-ph/0109120",
    archivePrefix = "arXiv",
    reportNumber = "WM-01-111, BU-01-22",
    doi = "10.1103/PhysRevD.65.013011",
    journal = "Phys. Rev. D",
    volume = "65",
    pages = "013011",
    year = "2002"
}

@article{Petcov:2018snn,
    author = "Petcov, S. T. and Titov, A. V.",
    title = "{Assessing the Viability of $A_4$, $S_4$ and $A_5$ Flavour Symmetries for Description of Neutrino Mixing}",
    eprint = "1804.00182",
    archivePrefix = "arXiv",
    primaryClass = "hep-ph",
    reportNumber = "SISSA 15/2018/FISI, IPMU18-0058, IPPP/18/22, SISSA-15-2018-FISI, IPPP-18-22",
    doi = "10.1103/PhysRevD.97.115045",
    journal = "Phys. Rev. D",
    volume = "97",
    number = "11",
    pages = "115045",
    year = "2018"
}

@article{Choi:2020rgn,
    author = "Choi, Kiwoon and Im, Sang Hui and Sub Shin, Chang",
    title = "{Recent Progress in the Physics of Axions and Axion-Like Particles}",
    eprint = "2012.05029",
    archivePrefix = "arXiv",
    primaryClass = "hep-ph",
    reportNumber = "CTPU-PTC-20-28",
    doi = "10.1146/annurev-nucl-120720-031147",
    journal = "Ann. Rev. Nucl. Part. Sci.",
    volume = "71",
    pages = "225--252",
    year = "2021"
}

@article{Chang:2021myh,
    author = "Chang, Chia-Hung Vincent and Chen, Chuan-Ren and Ho, Shu-Yu and Tseng, Shih-Yen",
    title = "{Explaining the MiniBooNE anomalous excess via a leptophilic ALP-sterile neutrino coupling}",
    eprint = "2102.05012",
    archivePrefix = "arXiv",
    primaryClass = "hep-ph",
    reportNumber = "KIAS-P21006",
    doi = "10.1103/PhysRevD.104.015030",
    journal = "Phys. Rev. D",
    volume = "104",
    number = "1",
    pages = "015030",
    year = "2021"
}

@article{Kim:2021eye,
    author = "Kim, Dongok and Kim, Younggeun and Semertzidis, Yannis K. and Shin, Yun Chang and Yin, Wen",
    title = "{Cosmic axion force}",
    eprint = "2105.03422",
    archivePrefix = "arXiv",
    primaryClass = "hep-ph",
    reportNumber = "TU-1023",
    doi = "10.1103/PhysRevD.104.095010",
    journal = "Phys. Rev. D",
    volume = "104",
    number = "9",
    pages = "095010",
    year = "2021"
}

@article{Ren:2021prq,
    author = "Ren, Jie and Wang, Daohan and Wu, Lei and Yang, Jin Min and Zhang, Mengchao",
    title = "{Detecting an axion-like particle with machine learning at the LHC}",
    eprint = "2106.07018",
    archivePrefix = "arXiv",
    primaryClass = "hep-ph",
    doi = "10.1007/JHEP11(2021)138",
    journal = "JHEP",
    volume = "11",
    pages = "138",
    year = "2021"
}

@article{Takahashi:2020uio,
    author = "Takahashi, Fuminobu and Yamada, Masaki and Yin, Wen",
    title = "{What if ALP dark matter for the XENON1T excess is the inflaton}",
    eprint = "2007.10311",
    archivePrefix = "arXiv",
    primaryClass = "hep-ph",
    reportNumber = "TU-1106, IPMU20-0081",
    doi = "10.1007/JHEP01(2021)152",
    journal = "JHEP",
    volume = "01",
    pages = "152",
    year = "2021"
}

@article{Jodidio:1986mz,
    author = "Jodidio, A. and others",
    title = "{Search for Right-Handed Currents in Muon Decay}",
    reportNumber = "LBL-21616",
    doi = "10.1103/PhysRevD.34.1967",
    journal = "Phys. Rev. D",
    volume = "34",
    pages = "1967",
    year = "1986",
    note = "[Erratum: Phys.Rev.D 37, 237 (1988)]"
}

@article{TWIST:2014ymv,
    author = "Bayes, R. and others",
    collaboration = "TWIST",
    title = "{Search for two body muon decay signals}",
    eprint = "1409.0638",
    archivePrefix = "arXiv",
    primaryClass = "hep-ex",
    doi = "10.1103/PhysRevD.91.052020",
    journal = "Phys. Rev. D",
    volume = "91",
    number = "5",
    pages = "052020",
    year = "2015"
}

@article{Bolton:1988af,
    author = "Bolton, R. D. and others",
    title = "{Search for Rare Muon Decays with the Crystal Box Detector}",
    reportNumber = "LA-UR-88-392",
    doi = "10.1103/PhysRevD.38.2077",
    journal = "Phys. Rev. D",
    volume = "38",
    pages = "2077",
    year = "1988"
}

@article{ARGUS:1995bjh,
    author = "Albrecht, H. and others",
    collaboration = "ARGUS",
    title = "{A Search for lepton flavor violating decays tau ----\ensuremath{>} e alpha, tau ---\ensuremath{>} mu alpha}",
    reportNumber = "DESY-95-071",
    doi = "10.1007/BF01579801",
    journal = "Z. Phys. C",
    volume = "68",
    pages = "25--28",
    year = "1995"
}

@phdthesis{Perrevoort:2018okj,
    author = "Perrevoort, Ann-Kathrin",
    title = "{Sensitivity Studies on New Physics in the Mu3e Experiment and Development of Firmware for the Front-End of the Mu3e Pixel Detector}",
    doi = "10.11588/heidok.00024585",
    school = "U. Heidelberg (main)",
    year = "2018"
}

@article{Griessinger:2017rpx,
    author = "Griessinger, Konrad",
    editor = "Yuan, Changzheng and Mo, Xiaohu and Wang, Liangliang",
    collaboration = "BaBar",
    title = "{New ISR Cross Section Results on $e^+e^- \to \pi^+\pi^-\pi^0\pi^0$ and $e^+e^- \to \pi^+\pi^-\eta$ from BaBar}",
    eprint = "1706.07678",
    archivePrefix = "arXiv",
    primaryClass = "hep-ex",
    doi = "10.1016/j.nuclphysbps.2017.03.042",
    journal = "Nucl. Part. Phys. Proc.",
    volume = "287-288",
    pages = "47--51",
    year = "2017"
}

@article{Bjorkeroth:2018dzu,
    author = {Bj\"orkeroth, Fredrik and Chun, Eung Jin and King, Stephen F.},
    title = "{Flavourful Axion Phenomenology}",
    eprint = "1806.00660",
    archivePrefix = "arXiv",
    primaryClass = "hep-ph",
    doi = "10.1007/JHEP08(2018)117",
    journal = "JHEP",
    volume = "08",
    pages = "117",
    year = "2018"
}

@article{Calibbi:2020jvd,
    author = "Calibbi, Lorenzo and Redigolo, Diego and Ziegler, Robert and Zupan, Jure",
    title = "{Looking forward to lepton-flavor-violating ALPs}",
    eprint = "2006.04795",
    archivePrefix = "arXiv",
    primaryClass = "hep-ph",
    reportNumber = "P3H-20-024, TTP20-025",
    doi = "10.1007/JHEP09(2021)173",
    journal = "JHEP",
    volume = "09",
    pages = "173",
    year = "2021"
}

@article{XENON:2020rca,
    author = "Aprile, E. and others",
    collaboration = "XENON",
    title = "{Excess electronic recoil events in XENON1T}",
    eprint = "2006.09721",
    archivePrefix = "arXiv",
    primaryClass = "hep-ex",
    doi = "10.1103/PhysRevD.102.072004",
    journal = "Phys. Rev. D",
    volume = "102",
    number = "7",
    pages = "072004",
    year = "2020"
}

@article{Takahashi:2020bpq,
    author = "Takahashi, Fuminobu and Yamada, Masaki and Yin, Wen",
    title = "{XENON1T Excess from Anomaly-Free Axionlike Dark Matter and Its Implications for Stellar Cooling Anomaly}",
    eprint = "2006.10035",
    archivePrefix = "arXiv",
    primaryClass = "hep-ph",
    reportNumber = "TU-1104, IPMU20-0069",
    doi = "10.1103/PhysRevLett.125.161801",
    journal = "Phys. Rev. Lett.",
    volume = "125",
    number = "16",
    pages = "161801",
    year = "2020"
}

@article{deSalas:2020pgw,
    author = "de~Salas, P. F. and Forero, D. V. and Gariazzo, S. and Mart{\'\i}nez-Mirav{\'e}, P. and Mena, O. and Ternes, C. A. and T{\'o}rtola, M. and Valle, J. W. F.",
    title = "{2020 global reassessment of the neutrino oscillation picture}",
    eprint = "2006.11237",
    archivePrefix = "arXiv",
    primaryClass = "hep-ph",
    doi = "10.1007/JHEP02(2021)071",
    journal = "JHEP",
    volume = "02",
    pages = "071",
    year = "2021"
}

@article{Esteban:2020cvm,
    author = "Esteban, Ivan and Gonzalez-Garcia, M. C. and Maltoni, Michele and Schwetz, Thomas and Zhou, Albert",
    title = "{The fate of hints: updated global analysis of three-flavor neutrino oscillations}",
    eprint = "2007.14792",
    archivePrefix = "arXiv",
    primaryClass = "hep-ph",
    reportNumber = "IFT-UAM/CSIC-112, YITP-SB-2020-21",
    doi = "10.1007/JHEP09(2020)178",
    journal = "JHEP",
    volume = "09",
    pages = "178",
    year = "2020"
}

@article{Fedele:2020fvh,
    author = "Fedele, Marco and Mastroddi, Alessio and Valli, Mauro",
    title = "{Minimal Froggatt-Nielsen Textures}",
    eprint = "2009.05587",
    archivePrefix = "arXiv",
    primaryClass = "hep-ph",
    reportNumber = "UCI-TR-2020-13",
    month = "9",
    year = "2020"
}

@article{Han:2020dwo,
    author = "Han, C. and L\'opez-Ib\'a\~nez, M. L. and Melis, A. and Vives, O. and Yang, J. M.",
    title = "{Anomaly-free leptophilic axionlike particle and its flavor violating tests}",
    eprint = "2007.08834",
    archivePrefix = "arXiv",
    primaryClass = "hep-ph",
    reportNumber = "FTUV-20-0717, IFIC/20-36",
    doi = "10.1103/PhysRevD.103.035028",
    journal = "Phys. Rev. D",
    volume = "103",
    number = "3",
    pages = "035028",
    year = "2021"
}

@article{Das:2016czs,
      author         = "Das, Dipankar and L\'opez-Ibáñez, M. L. and P\'erez, M.
                        Jay and Vives, Oscar",
      title          = "{Effective theories of flavor and the nonuniversal MSSM}",
      journal        = "Phys. Rev.",
      volume         = "D95",
      year           = "2017",
      number         = "3",
      pages          = "035001",
      doi            = "10.1103/PhysRevD.95.035001",
      eprint         = "1607.06827",
      archivePrefix  = "arXiv",
      primaryClass   = "hep-ph",
      reportNumber   = "FTUV-16-0719, IFIC-16-51",
      SLACcitation   = "%%CITATION = ARXIV:1607.06827;%%"
}

@article{Baldini:2018nnn,
      author         = "Baldini, A. M. and others",
      title          = "{The design of the MEG II experiment}",
      collaboration  = "MEG II",
      journal        = "Eur. Phys. J.",
      volume         = "C78",
      year           = "2018",
      number         = "5",
      pages          = "380",
      doi            = "10.1140/epjc/s10052-018-5845-6",
      eprint         = "1801.04688",
      archivePrefix  = "arXiv",
      primaryClass   = "physics.ins-det",
      SLACcitation   = "%%CITATION = ARXIV:1801.04688;%%"
}

@article{Froggatt:1978nt,
      author         = "Froggatt, C. D. and Nielsen, Holger Bech",
      title          = "{Hierarchy of Quark Masses, Cabibbo Angles and CP
                        Violation}",
      journal        = "Nucl. Phys.",
      volume         = "B147",
      year           = "1979",
      pages          = "277-298",
      doi            = "10.1016/0550-3213(79)90316-X",
      reportNumber   = "CERN-TH-2519",
      SLACcitation   = "%%CITATION = NUPHA,B147,277;%%"
}

@article{Leurer:1992wg,
      author         = "Leurer, Miriam and Nir, Yosef and Seiberg, Nathan",
      title          = "{Mass matrix models}",
      journal        = "Nucl. Phys.",
      volume         = "B398",
      year           = "1993",
      pages          = "319-342",
      doi            = "10.1016/0550-3213(93)90112-3",
      eprint         = "hep-ph/9212278",
      archivePrefix  = "arXiv",
      primaryClass   = "hep-ph",
      reportNumber   = "RU-92-59, WIS-92-94-PH",
      SLACcitation   = "%%CITATION = HEP-PH/9212278;%%"
}

@article{Leurer:1993gy,
      author         = "Leurer, Miriam and Nir, Yosef and Seiberg, Nathan",
      title          = "{Mass matrix models: The Sequel}",
      journal        = "Nucl. Phys.",
      volume         = "B420",
      year           = "1994",
      pages          = "468-504",
      doi            = "10.1016/0550-3213(94)90074-4",
      eprint         = "hep-ph/9310320",
      archivePrefix  = "arXiv",
      primaryClass   = "hep-ph",
      reportNumber   = "RU-93-43, WIS-93-93-PH",
      SLACcitation   = "%%CITATION = HEP-PH/9310320;%%"
}

@article{Nir:1993mx,
      author         = "Nir, Yosef and Seiberg, Nathan",
      title          = "{Should squarks be degenerate?}",
      journal        = "Phys. Lett.",
      volume         = "B309",
      year           = "1993",
      pages          = "337-343",
      doi            = "10.1016/0370-2693(93)90942-B",
      eprint         = "hep-ph/9304307",
      archivePrefix  = "arXiv",
      primaryClass   = "hep-ph",
      reportNumber   = "RU-93-16, WIS-93-37-PH",
      SLACcitation   = "%%CITATION = HEP-PH/9304307;%%"
}

@article{Dine:1993np,
      author         = "Dine, Michael and Leigh, Robert G. and Kagan, Alex",
      title          = "{Flavor symmetries and the problem of squark degeneracy}",
      journal        = "Phys. Rev.",
      volume         = "D48",
      year           = "1993",
      pages          = "4269-4274",
      doi            = "10.1103/PhysRevD.48.4269",
      eprint         = "hep-ph/9304299",
      archivePrefix  = "arXiv",
      primaryClass   = "hep-ph",
      reportNumber   = "SLAC-PUB-6147, SCIPP-93-04",
      SLACcitation   = "%%CITATION = HEP-PH/9304299;%%"
}

@article{Ibanez:1994ig,
      author         = "Ibanez, Luis E. and Ross, Graham G.",
      title          = "{Fermion masses and mixing angles from gauge symmetries}",
      journal        = "Phys. Lett.",
      volume         = "B332",
      year           = "1994",
      pages          = "100-110",
      doi            = "10.1016/0370-2693(94)90865-6",
      eprint         = "hep-ph/9403338",
      archivePrefix  = "arXiv",
      primaryClass   = "hep-ph",
      reportNumber   = "OUTP-94-03-P, FTUAM-94-7",
      SLACcitation   = "%%CITATION = HEP-PH/9403338;%%"
}

@article{Barbieri:1995uv,
      author         = "Barbieri, Riccardo and Dvali, G. R. and Hall, Lawrence
                        J.",
      title          = "{Predictions from a U(2) flavor symmetry in
                        supersymmetric theories}",
      journal        = "Phys. Lett.",
      volume         = "B377",
      year           = "1996",
      pages          = "76-82",
      doi            = "10.1016/0370-2693(96)00318-8",
      eprint         = "hep-ph/9512388",
      archivePrefix  = "arXiv",
      primaryClass   = "hep-ph",
      reportNumber   = "LBL-38065, UCB-PTH-95-44",
      SLACcitation   = "%%CITATION = HEP-PH/9512388;%%"
}

@article{Pomarol:1995xc,
      author         = "Pomarol, Alex and Tommasini, Daniele",
      title          = "{Horizontal symmetries for the supersymmetric flavor
                        problem}",
      journal        = "Nucl. Phys.",
      volume         = "B466",
      year           = "1996",
      pages          = "3-24",
      doi            = "10.1016/0550-3213(96)00074-0",
      eprint         = "hep-ph/9507462",
      archivePrefix  = "arXiv",
      primaryClass   = "hep-ph",
      reportNumber   = "CERN-TH-95-207",
      SLACcitation   = "%%CITATION = HEP-PH/9507462;%%"
}

@article{Dudas:1995yu,
      author         = "Dudas, E. and Pokorski, S. and Savoy, Carlos A.",
      title          = "{Yukawa matrices from a spontaneously broken Abelian
                        symmetry}",
      journal        = "Phys. Lett.",
      volume         = "B356",
      year           = "1995",
      pages          = "45-55",
      doi            = "10.1016/0370-2693(95)00795-M",
      eprint         = "hep-ph/9504292",
      archivePrefix  = "arXiv",
      primaryClass   = "hep-ph",
      reportNumber   = "SACLAY-SPH-T-95-027, MPI-PTH-95-33",
      SLACcitation   = "%%CITATION = HEP-PH/9504292;%%"
}

@article{Binetruy:1996cs,
      author         = "Binetruy, Pierre and Lavignac, Stephane and Petcov,
                        Serguey T. and Ramond, Pierre",
      title          = "{Quasidegenerate neutrinos from an Abelian family
                        symmetry}",
      journal        = "Nucl. Phys.",
      volume         = "B496",
      year           = "1997",
      pages          = "3-23",
      doi            = "10.1016/S0550-3213(97)00211-3",
      eprint         = "hep-ph/9610481",
      archivePrefix  = "arXiv",
      primaryClass   = "hep-ph",
      reportNumber   = "LPTHE-ORSAY-96-63, UFIFT-HEP-96-25, SISSA-145-96-EP",
      SLACcitation   = "%%CITATION = HEP-PH/9610481;%%"
}

@article{Barbieri:1996ww,
      author         = "Barbieri, Riccardo and Hall, Lawrence J. and Raby, Stuart
                        and Romanino, Andrea",
      title          = "{Unified theories with U(2) flavor symmetry}",
      journal        = "Nucl. Phys.",
      volume         = "B493",
      year           = "1997",
      pages          = "3-26",
      doi            = "10.1016/S0550-3213(97)00134-X",
      eprint         = "hep-ph/9610449",
      archivePrefix  = "arXiv",
      primaryClass   = "hep-ph",
      reportNumber   = "IFUP-TH-61-96, LBL-39488, OHSTPY-HEP-T-96-033,
                        UCB-PTH-96-45",
      SLACcitation   = "%%CITATION = HEP-PH/9610449;%%"
}

@article{Choi:1998wc,
      author         = "Choi, Kiwoon and Hwang, Kyuwan and Chun, Eung Jin",
      title          = "{Atmospheric and solar neutrino masses from horizontal
                        U(1) symmetry}",
      journal        = "Phys. Rev.",
      volume         = "D60",
      year           = "1999",
      pages          = "031301",
      doi            = "10.1103/PhysRevD.60.031301",
      eprint         = "hep-ph/9811363",
      archivePrefix  = "arXiv",
      primaryClass   = "hep-ph",
      reportNumber   = "KAIST-TH-98-19A, KAIST-TH-98-19, KIAS-P98040",
      SLACcitation   = "%%CITATION = HEP-PH/9811363;%%"
}

@article{Barbieri:1999km,
      author         = "Barbieri, Riccardo and Hall, Lawrence J. and Kane, Gordon
                        L. and Ross, Graham G.",
      title          = "{Nearly degenerate neutrinos and broken flavor symmetry}",
      year           = "1999",
      eprint         = "hep-ph/9901228",
      archivePrefix  = "arXiv",
      primaryClass   = "hep-ph",
      reportNumber   = "OUTP-9901-P, UCB-PTH-98-57, LBNL-42572, SNS-PH-98-25,
                        LBL-42572",
      SLACcitation   = "%%CITATION = HEP-PH/9901228;%%"
}

@article{King:2001uz,
      author         = "King, S. F. and Ross, Graham G.",
      title          = "{Fermion masses and mixing angles from SU(3) family
                        symmetry}",
      journal        = "Phys. Lett.",
      volume         = "B520",
      year           = "2001",
      pages          = "243-253",
      doi            = "10.1016/S0370-2693(01)01139-X",
      eprint         = "hep-ph/0108112",
      archivePrefix  = "arXiv",
      primaryClass   = "hep-ph",
      reportNumber   = "SHEP-01-21, OUTP-01-46P",
      SLACcitation   = "%%CITATION = HEP-PH/0108112;%%"
}

@article{Babu:2002dz,
      author         = "Babu, K. S. and Ma, Ernest and Valle, J. W. F.",
      title          = "{Underlying A(4) symmetry for the neutrino mass matrix
                        and the quark mixing matrix}",
      journal        = "Phys. Lett.",
      volume         = "B552",
      year           = "2003",
      pages          = "207-213",
      doi            = "10.1016/S0370-2693(02)03153-2",
      eprint         = "hep-ph/0206292",
      archivePrefix  = "arXiv",
      primaryClass   = "hep-ph",
      reportNumber   = "UCRHEP-T341, OSU-HEP-02-07, IFIC-02-26",
      SLACcitation   = "%%CITATION = HEP-PH/0206292;%%"
}

@article{Altarelli:2002sg,
      author         = "Altarelli, Guido and Feruglio, Ferruccio and Masina,
                        Isabella",
      title          = "{Models of neutrino masses: Anarchy versus hierarchy}",
      journal        = "JHEP",
      volume         = "01",
      year           = "2003",
      pages          = "035",
      doi            = "10.1088/1126-6708/2003/01/035",
      eprint         = "hep-ph/0210342",
      archivePrefix  = "arXiv",
      primaryClass   = "hep-ph",
      reportNumber   = "DFPD-02-TH-23, CERN-TH-2002-250, SACLAY-T02-138",
      SLACcitation   = "%%CITATION = HEP-PH/0210342;%%"
}

@article{Ross:2004qn,
      author         = "Ross, Graham G. and Velasco-Sevilla, Liliana and Vives,
                        Oscar",
      title          = "{Spontaneous CP violation and nonAbelian family symmetry
                        in SUSY}",
      journal        = "Nucl. Phys.",
      volume         = "B692",
      year           = "2004",
      pages          = "50-82",
      doi            = "10.1016/j.nuclphysb.2004.05.020",
      eprint         = "hep-ph/0401064",
      archivePrefix  = "arXiv",
      primaryClass   = "hep-ph",
      reportNumber   = "OUTP-0402P, MCTP-03-61, FTUV-04-0110",
      SLACcitation   = "%%CITATION = HEP-PH/0401064;%%"
}

@article{Chankowski:2005qp,
      author         = "Chankowski, Piotr H. and Kowalska, Kamila and Lavignac,
                        Stephane and Pokorski, Stefan",
      title          = "{Update on fermion mass models with an anomalous
                        horizontal U(1) symmetry}",
      journal        = "Phys. Rev.",
      volume         = "D71",
      year           = "2005",
      pages          = "055004",
      doi            = "10.1103/PhysRevD.71.055004",
      eprint         = "hep-ph/0501071",
      archivePrefix  = "arXiv",
      primaryClass   = "hep-ph",
      reportNumber   = "IFT-2004-31, CERN-PH-TH-2004-263",
      SLACcitation   = "%%CITATION = HEP-PH/0501071;%%"
}

@article{deMedeirosVarzielas:2005qg,
      author         = "de Medeiros Varzielas, I. and King, S. F. and Ross, G.
                        G.",
      title          = "{Tri-bimaximal neutrino mixing from discrete subgroups of
                        SU(3) and SO(3) family symmetry}",
      journal        = "Phys. Lett.",
      volume         = "B644",
      year           = "2007",
      pages          = "153-157",
      doi            = "10.1016/j.physletb.2006.11.015",
      eprint         = "hep-ph/0512313",
      archivePrefix  = "arXiv",
      primaryClass   = "hep-ph",
      SLACcitation   = "%%CITATION = HEP-PH/0512313;%%"
}

@article{Altarelli:2005yx,
      author         = "Altarelli, Guido and Feruglio, Ferruccio",
      title          = "{Tri-bimaximal neutrino mixing, A(4) and the modular
                        symmetry}",
      journal        = "Nucl. Phys.",
      volume         = "B741",
      year           = "2006",
      pages          = "215-235",
      doi            = "10.1016/j.nuclphysb.2006.02.015",
      eprint         = "hep-ph/0512103",
      archivePrefix  = "arXiv",
      primaryClass   = "hep-ph",
      reportNumber   = "CERN-PH-TH-2005-226",
      SLACcitation   = "%%CITATION = HEP-PH/0512103;%%"
}

@article{Luhn:2007sy,
      author         = "Luhn, Christoph and Nasri, Salah and Ramond, Pierre",
      title          = "{Tri-bimaximal neutrino mixing and the family symmetry
                        semidirect product of Z(7) and Z(3)}",
      journal        = "Phys. Lett.",
      volume         = "B652",
      year           = "2007",
      pages          = "27-33",
      doi            = "10.1016/j.physletb.2007.06.059",
      eprint         = "0706.2341",
      archivePrefix  = "arXiv",
      primaryClass   = "hep-ph",
      reportNumber   = "UFIFT-HEP-07-8",
      SLACcitation   = "%%CITATION = ARXIV:0706.2341;%%"
}

@article{Babu:2011mv,
      author         = "Babu, K. S. and Kawashima, Kenji and Kubo, Jisuke",
      title          = "{Variations on the Supersymmetric $Q_6$ Model of Flavor}",
      journal        = "Phys. Rev.",
      volume         = "D83",
      year           = "2011",
      pages          = "095008",
      doi            = "10.1103/PhysRevD.83.095008",
      eprint         = "1103.1664",
      archivePrefix  = "arXiv",
      primaryClass   = "hep-ph",
      reportNumber   = "OSU-HEP-11-03, KANAZAWA-11-05",
      SLACcitation   = "%%CITATION = ARXIV:1103.1664;%%"
}

@article{Chen:2014wiw,
      author         = "Chen, Gaoli and P\'erez, M. Jay and Ramond, Pierre",
      title          = "{Neutrino masses, the $\mu$-term and $\mathcal{
                        PSL}_2(7)$}",
      journal        = "Phys. Rev.",
      volume         = "D92",
      year           = "2015",
      number         = "7",
      pages          = "076006",
      doi            = "10.1103/PhysRevD.92.076006",
      eprint         = "1412.6107",
      archivePrefix  = "arXiv",
      primaryClass   = "hep-ph",
      SLACcitation   = "%%CITATION = ARXIV:1412.6107;%%"
}

@article{deMedeirosVarzielas:2005ax,
      author         = "de Medeiros Varzielas, Ivo and Ross, Graham G.",
      title          = "{SU(3) family symmetry and neutrino bi-tri-maximal
                        mixing}",
      journal        = "Nucl. Phys.",
      volume         = "B733",
      year           = "2006",
      pages          = "31-47",
      doi            = "10.1016/j.nuclphysb.2005.10.039",
      eprint         = "hep-ph/0507176",
      archivePrefix  = "arXiv",
      primaryClass   = "hep-ph",
      reportNumber   = "OUTP-0420P",
      SLACcitation   = "%%CITATION = HEP-PH/0507176;%%"
}

@article{Altarelli:2005yp,
      author         = "Altarelli, Guido and Feruglio, Ferruccio",
      title          = "{Tri-bimaximal neutrino mixing from discrete symmetry in
                        extra dimensions}",
      journal        = "Nucl. Phys.",
      volume         = "B720",
      year           = "2005",
      pages          = "64-88",
      doi            = "10.1016/j.nuclphysb.2005.05.005",
      eprint         = "hep-ph/0504165",
      archivePrefix  = "arXiv",
      primaryClass   = "hep-ph",
      reportNumber   = "DFPD-05-TH-14, CERN-PH-TH-2005-067",
      SLACcitation   = "%%CITATION = HEP-PH/0504165;%%"
}

@article{Altarelli:2006kg,
      author         = "Altarelli, Guido and Feruglio, Ferruccio and Lin, Yin",
      title          = "{Tri-bimaximal neutrino mixing from orbifolding}",
      journal        = "Nucl. Phys.",
      volume         = "B775",
      year           = "2007",
      pages          = "31-44",
      doi            = "10.1016/j.nuclphysb.2007.03.042",
      eprint         = "hep-ph/0610165",
      archivePrefix  = "arXiv",
      primaryClass   = "hep-ph",
      reportNumber   = "DFPD-06-TH-12, RM3-TH-06-17, CERN-PH-TH-2006-205",
      SLACcitation   = "%%CITATION = HEP-PH/0610165;%%"
}

@article{Ma:1991eg,
      author         = "Ma, Ernest",
      title          = "{S(3) Z(3) model of lepton mass matrices}",
      journal        = "Phys. Rev.",
      volume         = "D44",
      year           = "1991",
      pages          = "587-589",
      doi            = "10.1103/PhysRevD.44.587",
      reportNumber   = "UCRHEP-T68",
      SLACcitation   = "%%CITATION = PHRVA,D44,587;%%"
}

@article{Ding:2011cm,
      author         = "Ding, Gui-Jun and Everett, Lisa L. and Stuart, Alexander
                        J.",
      title          = "{Golden Ratio Neutrino Mixing and $A_5$ Flavor Symmetry}",
      journal        = "Nucl. Phys.",
      volume         = "B857",
      year           = "2012",
      pages          = "219-253",
      doi            = "10.1016/j.nuclphysb.2011.12.004",
      eprint         = "1110.1688",
      archivePrefix  = "arXiv",
      primaryClass   = "hep-ph",
      reportNumber   = "MADPH-11-1576",
      SLACcitation   = "%%CITATION = ARXIV:1110.1688;%%"
}

@article{Lopez-Ibanez:2017xxw,
      author         = "López-Ibáñez, M. Luisa and Melis, Aurora and Pérez,
                        M. Jay and Vives, Oscar",
      title          = "{Slepton non-universality in the flavor-effective MSSM}",
      journal        = "JHEP",
      volume         = "11",
      year           = "2017",
      pages          = "162",
      doi            = "10.1007/JHEP11(2017)162, 10.1007/JHEP04(2018)015",
      note           = "[Erratum: JHEP04,015(2018)]",
      eprint         = "1710.02593",
      archivePrefix  = "arXiv",
      primaryClass   = "hep-ph",
      SLACcitation   = "%%CITATION = ARXIV:1710.02593;%%"
}

@article{deMedeirosVarzielas:2017sdv,
      author         = "de Medeiros Varzielas, Ivo and Ross, Graham G. and
                        Talbert, Jim",
      title          = "{A Unified Model of Quarks and Leptons with a Universal
                        Texture Zero}",
      year           = "2017",
      eprint         = "1710.01741",
      archivePrefix  = "arXiv",
      primaryClass   = "hep-ph",
      reportNumber   = "OUTP-17-13P, DESY-17-146",
      SLACcitation   = "%%CITATION = ARXIV:1710.01741;%%"
}

@article{Lopez-Ibanez:2019rgb,
      author         = "López-Ibáñez, M. L. and Melis, Aurora and Meloni,
                        Davide and Vives, Oscar",
      title          = "{Lepton flavor violation and neutrino masses from A$_{5}$
                        and CP in the non-universal MSSM}",
      journal        = "JHEP",
      volume         = "06",
      year           = "2019",
      pages          = "047",
      doi            = "10.1007/JHEP06(2019)047",
      eprint         = "1901.04526",
      archivePrefix  = "arXiv",
      primaryClass   = "hep-ph",
      SLACcitation   = "%%CITATION = ARXIV:1901.04526;%%"
}

\end{document}